\documentclass[twocolumn]{aastex6}

\usepackage{amsmath}
\usepackage[T1]{fontenc}

\newcommand{\Asixone}{ALESS\,61.1}
\newcommand{\Asixfive}{ALESS\,65.1}
\newcommand{\Aseventhree}{ALESS\,73.1}
\newcommand{\Afourseven}{UDS\,47.0}
\newcommand{\CII}{[C{\small II}]}
\newcommand{\LCII}{$L_{\text{[CII]}}$}
\newcommand{\LFIR}{$L_{\text{FIR}}$}
\newcommand{\densfr}{$\Sigma_{\text{SFR}}$}
\newcommand{\cont}{345\,\text{GHz}}
\newcommand{\sersic}{S\'ersic}

\fullcollaborationName{The Friends of AASTeX Collaboration}

\begin{document}


\title{The dust and \CII\ morphologies of redshift $\sim4.5$ sub-millimeter galaxies at $\sim200$\,pc resolution: The absence of large clumps in the interstellar medium of high-redshift galaxies}

\author{
B. Gullberg\altaffilmark{1}, 
A. M. Swinbank\altaffilmark{1}, 
I. Smail\altaffilmark{1},  
A. D. Biggs\altaffilmark{2},
F. Bertoldi\altaffilmark{3},
C. De Breuck\altaffilmark{2},
S. C. Chapman\altaffilmark{4}, 
C.-C. Chen\altaffilmark{2}, 
E. A. Cooke\altaffilmark{1}, 
K. E. K. Coppin\altaffilmark{5},
P. Cox\altaffilmark{6},
H. Dannerbauer\altaffilmark{7},
J. S. Dunlop\altaffilmark{8}, 
A. C. Edge\altaffilmark{1}, 
D. Farrah\altaffilmark{9}, 
J. E. Geach\altaffilmark{6}, 
T. R. Greve\altaffilmark{10}, 
J. Hodge\altaffilmark{11}, 
E. Ibar\altaffilmark{12}, 
R. J. Ivison\altaffilmark{2,8},
A. Karim\altaffilmark{13}, 
E. Schinnerer\altaffilmark{14}, 
D. Scott\altaffilmark{15}, 
J. M. Simpson\altaffilmark{16}, 
S. M. Stach\altaffilmark{1}, 
A. P. Thomson\altaffilmark{1,17}, 
P. van der Werf\altaffilmark{11}, 
F. Walter\altaffilmark{14}, 
J. L. Wardlow\altaffilmark{1} \&
A. Weiss\altaffilmark{18}
}

\altaffiltext{1}{Centre for Extragalactic Astronomy, Department of Physics, Durham University, South Road, Durham DH1 3LE, UK}
\altaffiltext{2}{European Southern Observatory, Karl-Schwarzschild-Stra\ss e 2, D-85748 Garching bei M\"unchen, Germany}
\altaffiltext{3}{Instituto de Astrofísica de Canarias (IAC), E-38205 La Laguna, Tenerife, Spain}
\altaffiltext{4}{Department of Physics and Atmospheric Science, Dalhousie University, Halifax, Canada}
\altaffiltext{5}{Centre for Astrophysics Research, Science \& Technology Research Institute, University of Hertfordshire, Hatfield AL10 9AB, UK}
\altaffiltext{6}{Joint ALMA Observatory - ESO, Av. Alonso de Cordova, 3104, Santiago, Chile}
\altaffiltext{7}{Dpto. Astrofísica, Universidad de La Laguna, E-38206 La Laguna, Tenerife, Spain}
\altaffiltext{8}{Institute for Astronomy, University of Edinburgh, Royal Observatory, Blackford Hill, Edinburgh EH9 3HJ, UK}
\altaffiltext{9}{Department of Physics, Virginia Tech, Blacksburg, VA 24061, USA}
\altaffiltext{10}{University College London, Gower Street, London WC1E 6BT, UK}
\altaffiltext{11}{Leiden Observatory, Leiden University, P.O. Box 9513, 2300 RA Leiden, Netherlands}
\altaffiltext{12}{Instituto de Física y Astronomía, Universidad de Valparaíso, Avda. Gran Breta\~na 1111, 2340000 Valparaíso, Chile}
\altaffiltext{13}{Bonn University, Auf dem H\"ugel 71, D53121 Bonn, Germany}
\altaffiltext{14}{Max-Planck-Institut f\"ur Astronomie, K\"onigstuhl 17, D-69117, Heidelberg, Germany}
\altaffiltext{15}{Department of Physics \& Astronomy, University of British Columbia, 6224 Agricultural Road, Vancouver, BC V6T 1Z1, Canada}
\altaffiltext{16}{Academia Sinica Institute of Astronomy and Astrophysics, No. 1, Sec. 4, Roosevelt Road, Taipei 10617, Taiwan}
\altaffiltext{17}{Jodrell Bank Centre for Astrophysics, The University of Manchester, Oxford Road, Manchester, M13 9PL, UK}
\altaffiltext{18}{Max-Planck-Institut f\"ur Radioastronomie, Auf dem H\"ugel 69 D-53121 Bonn, Germany}
\altaffiltext{$^*$}{bitten.gullberg@durham.ac.uk}

\begin{abstract}
We present deep high resolution (0\farcs03, 200\,pc) ALMA Band 7 observations covering the dust continuum and \CII\ $\lambda157.7$\,$\mu$m emission in four $z\sim4.4-4.8$ sub-millimeter galaxies (SMGs) selected from the ALESS and AS2UDS surveys. The data show that the rest-frame 160\,$\mu$m (observed 345\,GHz) dust emission is consistent with smooth morphologies on kpc scales for three of the sources. One source, \Afourseven, displays apparent substructure but this is also consistent with a smooth morphology, as indicated by simulations showing that smooth exponential disks can appear clumpy when observed at high angular resolution (0\farcs03) and depth of these observations ($\sigma_{\cont} \sim27-47$\,$\mu$Jy\,beam$^{-1}$). The four SMGs are bright \CII\ emitters, and we extract \CII\ spectra from the high resolution data, and recover $\sim20-100$\% of the \CII\ flux and $\sim40-80$\% of the dust continuum emission, compared to the previous lower resolution observations. When tapered to 0\farcs2 resolution our maps recover $\sim80-100$\% of the continuum emission, indicating that $\sim60$\% of the emission is resolved out on $\sim200$\,pc scales. We find that the \CII\ emission in high-redshift galaxies is more spatially extended than the rest-frame 160\,$\mu$m dust continuum by a factor of $1.6\pm0.4$.  By considering the \LCII/\LFIR\ ratio as a function of the star-formation rate surface density (\densfr) we revisit the \CII\ deficit, and suggest that the decline in the \LCII/\LFIR\ ratio as a function of \densfr\ is consistent with local processes. We also explore the physical drivers that may be responsible for these trends and can give rise to the properties found in the densest regions of SMGs. 

\end{abstract}
\keywords{galaxies: evolution -- submillimeter: galaxies -- galaxies: ISM}

\section{Introduction} \label{sec:intro}
\noindent
The most luminous galaxies at high redshift ($z>1$) are dusty star-forming galaxies, originally identified at sub-millimeter wavelengths and therefore known as sub-millimeter galaxies \citep[SMGs,][]{casey14}.
This galaxy population has many properties similar to those of local ultra-luminous galaxies (ULIRGs, \citealt{sanders96}), such as high infrared luminosities (typically $L_{\text{FIR}}>10^{12}$\,L$_{\odot}$), as well as high gas and dynamical masses and gas fractions \citep[e.g.,][]{tacconi08, engel10, riechers11, bothwell13}. 
However, studies have shown important differences between SMGs and ULIRGs. 
For example, the spatial extent of the gas and star formation in SMGs appears to be much larger than that typically seen in local ULIRGs ($\sim$\ few kpc in SMGs compared to just 100s of pc in local ULIRGs, e.g., \citealt{chapman04, sakamoto08, kennicutt11, ivison12, croxall12, simpson15b, ikarashi15, hodge16}), and while the intense star formation seen in local ULIRGs appears to be triggered by major mergers \citep[e.g.,][]{clements96, farrah01,surace01, veilleux02}, theoretical predictions have suggested that SMGs at $z\sim1-5$ comprise a heterogeneous mix of star formation occurring in extended disks, pre-coalescence mergers, and late-stage mergers \cite[e.g.,][]{hayward11, cowley17}, which may be consistent with \textit{Hubble Space Telescope} (\textit{HST}) imaging \citep{chen15}.

Rest-frame ultraviolet (UV)/optical observations of high-redshift `main-sequence' star-forming galaxies show `clumpy' star-forming structures, more massive and brighter than seen locally \citep[e.g.,][]{livermore12, genzel12}.
In a simple framework for gas collapse in a gas-rich disk the masses of these `clumps' are governed by the average gas surface density of the surrounding interstellar medium. 
In high-redshift galaxies with high gas fractions the masses of collapsing clouds are therefore expected to be shifted to higher masses. 
This could result in $10^8-10^9$\,M$_{\odot}$ `clumps' \citep[e.g.,][]{genzel12, forster-schreiber11, livermore12}, and since more massive regions host proportionally more star formation \citep{kennicutt88}, these giant clumps can dominate the galaxy morphology and so explain the clumpy nature of the UV/optical images of high-redshift galaxies \citep[e.g.,][]{elmegreen09, shibuya15}.

\begin{table*}
\centering          
\begin{tabular}{l  | c c c c c c c c c}
\hline\hline       
Source name         & $z$ &       R.A.              &      Dec.\          & Discovery & $S_{\cont}$ & $SdV_{\text{[CII]}}$  & FWHM$_{\text{[CII]}}$ & $L_{\text{FIR}}$\\
                              &        &  \multicolumn{2}{c}{(J2000)}      &   resolution  &         [mJy]    & [Jy\,km\,s$^{-1}$]   & [km\,s$^{-1}$]              & [$10^{12}$\,L$_{\odot}$]\\ 
\hline                    
\Asixone               & $4.4189\pm0.0004^a$ & 03:32:45.88 & $-$28:00:23.4 & $1\farcs8\times1\farcs2$     & $4.3\pm0.5$ & $2.5\pm0.4$ & $230\pm25$ & $3.1\pm0.2$ \\ 
\Asixfive               & $4.4445\pm0.0005^a$ & 03:32:52.25 & $-$27:35:26.2 & $1\farcs8\times1\farcs2$     & $4.2\pm0.4$ & $5.4\pm0.7$ & $490\pm35$ & $3.1\pm0.2$ \\ 
\Aseventhree       & $4.7555\pm0.0001^b$ & 03:32:29.30 & $-$27:56:19.6 & $0\farcs65\times0\farcs40$ & $6.6\pm0.2$ & $7.4\pm0.4$ &$375\pm105$ & $2.9\pm0.2$ \\ 
\Afourseven$^*$  & $4.4201\pm0.0001^c$  & 02:19:24.85 & $-$05:09:20.8 & $0\farcs35\times0\farcs25$ & $8.7\pm0.6$ & $4.3\pm0.9$ & $935\pm250$ & $3.2\pm0.4$ \\ 
\hline
\end{tabular}
\caption{{\small \textit{Column 1:} source names. The asterisk indicates that \Afourseven\ from \cite{simpson17} has since changed name to AS2UDS0051.0 in Stach et al. (in prep.). 
\textit{Column 2:} spectroscopic redshift from the observed \CII\ frequency, $a$) \cite{swinbank12}, $b$) \cite{debreuck14} and $c$) this work. 
\textit{Columns 3 \& 4:} source positions. 
\textit{Column 5:} resolution of the observations from ALMA Cycle 0 and 1. 
\textit{Column 6:} \cont\ dust continuum flux density from the lower resolution ALMA Cycle 0 and 1 observations \citep{swinbank12,debreuck14,simpson15b}. 
\textit{Column 7:} velocity integrated line flux of the \CII\ emission lines detected in ALMA Cycle 0 and 1 \citep{swinbank12,debreuck14}.
\textit{Column 8:} FWHM of the \CII\ emission lines detected in ALMA Cycle 0 and 1 \citep{swinbank12,debreuck14}.
\textit{Column 9:} infrared luminosity determined by assuming $T_{\text{d}}=50$\,K, as determined for 13 $z\sim4.4$ \CII\ identified SMGs in the UDS (Cooke et al. in prep.).}}
\label{table:cont_cy0}
\end{table*}

The resolution provided by ALMA is now allowing sub-millimeter observations on spatial scales comparable to those provided by optical and UV observations from \textit{HST}. 
Recent studies have searched for giant clumps at sub-millimeter wavelengths \citep[e.g.,][]{swinbank10, swinbank15, almapartnership15, iono16, oteo17}. 
In a study of 16 ALESS SMGs at 0\farcs16 resolution ($\sim1$\,kpc), \cite{hodge16} identified disk-like morphologies with no significant evidence for clumps in dust emission in the majority of their galaxies. 
However, giant H{\small II} regions in local galaxies are a few $100$\,pc \citep[e.g.,][]{hill05,sakamoto08}.
This means that, although this study measures structures on $\sim1$\,kpc scales, even higher resolution is required to search for extended clumpy disks with 200--500\,pc size clumps, as seen in some simulations \citep{dekel09,bournaud14} and locally \citep[e.g.,][]{hill05,sakamoto08}. 

One particularly powerful tool to study the structure of high-redshift galaxies in the sub-millimeter waveband is the bright \CII\ $\lambda157.7$\,$\mu$m line. 
This far-infrared (FIR) fine-structure emission line is emitted by the $^2$P$_{3/2}-^2$P$_{1/2}$ transition in singly ionized carbon (\CII) and accounts for up to $\sim1$\% of the cooling in the interstellar medium \citep{stacey91, brauher08, gracia_carpio11}. 
It is therefore one of the brightest and best studied atomic lines. 
The \CII\ emission line arises from both photo dominated regions (PDRs), which form on the UV-illuminated surfaces of molecular clouds, diffuse H{\small II} regions, and also from diffuse ISM \citep{madden93, lord96}. 
Early studies of \CII\ in local ULIRGs using the \textit{Kuiper Airborne Observatory} and \textit{Infrared Space Observatory} (\text{ISO}) \citep{stacey91, malhotra97, luhman98, malhotra01, luhman03} revealed a deficit in the \CII\ line strength compared to the far-infrared emission for lower luminosity galaxies.  
For galaxies with \LFIR$<10^{11}$\,L$_{\odot}$, the \LCII/\LFIR\ ratio is constant at $\sim1$\%, however at \LFIR$>10^{11}$\,L$_{\odot}$, the \LCII/\LFIR\ ratio decreases to $\sim0.1-0.01$\%. 
This decrease is known as the `\CII\ deficit', and many attempts have been made over the past two decades to investigate its origin \citep[e.g.,][]{hailey-dunsheath10, ivison10, stacey10, valtchanov11,gracia_carpio11, farrah13, gullberg15, spilker16, diaz-santos17}. 

Among the various explanations proposed for this behaviour are: \CII\ self-absorption, strong continuum extinction at 158\,$\mu$m; collisional quenching of \CII\ emission; high ionization parameters; and metallicity dependence (see \citealt{smith17} for an extensive discussion). 

By exploring the \LCII/\LFIR\ ratio as a function of the star-formation rate surface density (\densfr) in spatially resolved local galaxies in the KINGFISH sample, \cite{smith17} identify a declining relation of the \LCII/\LFIR\ ratio as a function of \densfr. The authors suggest that the \CII\ deficit is driven by local physical processes of interstellar gas (e.g., \CII\ self-absorption, dust extinction, and dust grain charge), not related to the global properties of the galaxies. Another study of spatially resolved local galaxies in the GOALS sample by \cite{diaz-santos17}, likewise suggest that local processes are the cause of the \CII\ deficit, proposing that the radiation field strength to gas density ratio is the driver. 

In this paper we present high-resolution (0\farcs03) ALMA Cycle 3 Band 7 observations of four SMGs at $z\sim4.4-4.8$, mapping their structure in dust and \CII\ emission on $\sim200$\,pc scales. 
Our observations show a range of morphologies in the observed \cont\ dust continuum emission (rest-frame 160\,$\mu$m) and \CII\ emission lines. 
In \S~\ref{sec:obs} we describe the observations and data reduction, while in \S~\ref{sec:analysis} we present our analysis and in \S~\ref{sec:disc} and \S~\ref{sec:con} our discussion and conclusions.
We assume a cosmology with $\Omega_{\Lambda} = 0.73$,  $\Omega_{\text{m}} = 0.27$ and $H_0 = 72$\,km\,s$^{-1}$\,Mpc$^{-1}$, in which 1\arcsec\ corresponds to a physical scale of 6.7\,kpc at $z\sim4.4$.

\section{Sample}
\noindent
Three of our targets (\Asixone, \Asixfive, and \Aseventhree) were selected from ALMA Band 7 (observed 870\,$\mu$m/\cont ) follow-up observations of sources detected in the single-dish LABOCA Extended \textit{Chandra} Deep Field South (ECDFS) Submm Survey (LESS, \citealt{weiss09}). 
The ALMA Cycle 0 continuum observations of these SMGs were reported in \cite{hodge13} (see Table~\ref{table:cont_cy0}) and revealed serendipitous detections of \CII\ in \Asixone\ and \Asixfive, establishing the redshifts as $z=4.4189$ and $z=4.4445$ respectively \citep{swinbank12}. 
The redshift of \Aseventhree\ was already known ($z=4.756$) and is also detected in \CII\ emission from ALMA Cycle 0 observation \citep{coppin09, debreuck14}.

\begin{figure*}
\centering
\includegraphics[trim=0.3cm 0.3cm 0.3cm 0.3cm, clip=true,scale=0.61,angle=90]{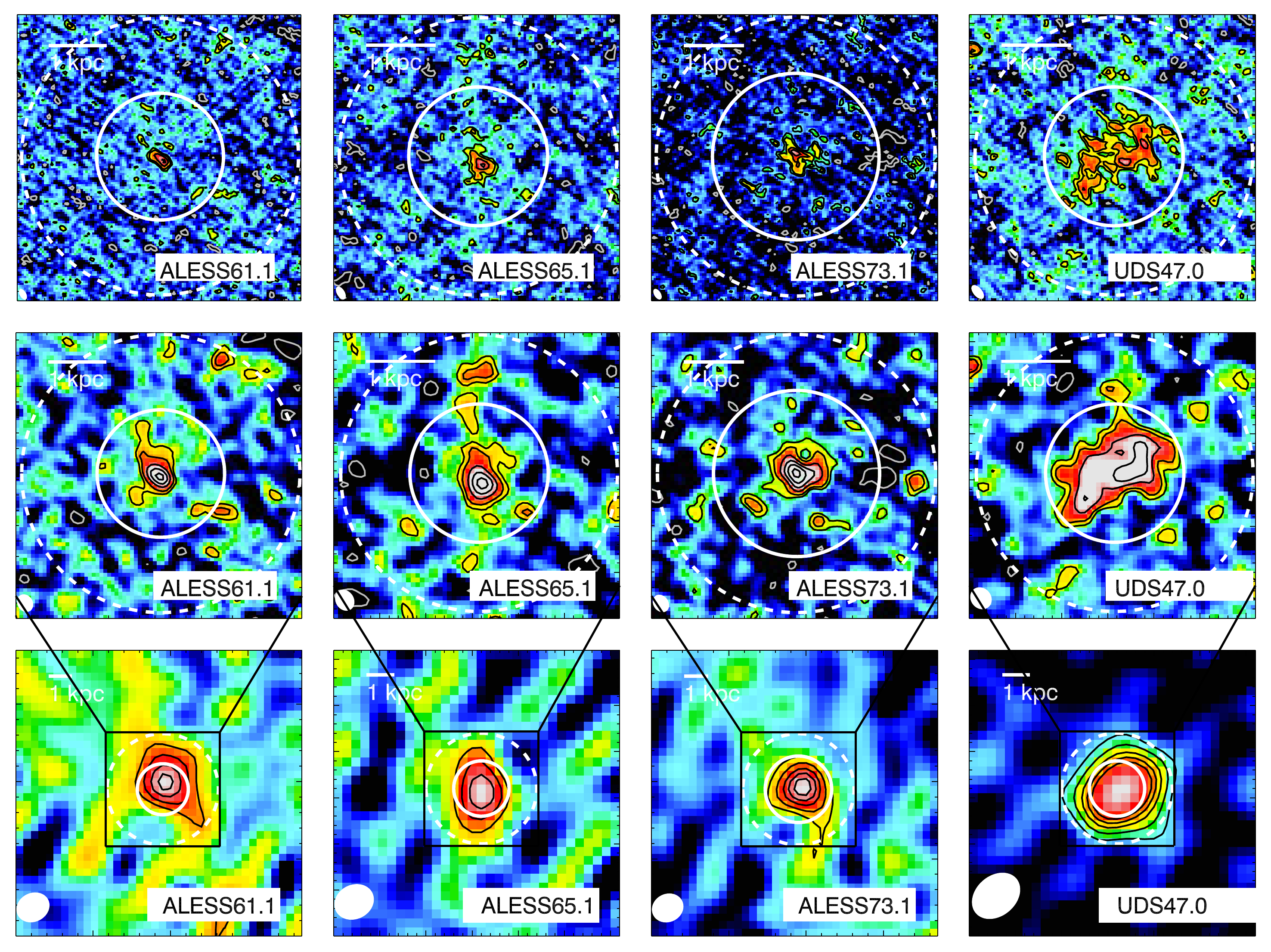}
\caption{{\small
Continuum maps at three different resolutions for our sample of SMGs. The white solid and white dashed circles indicate the sizes estimated for the \cont\ dust continuum emission and the \CII\ emitting gas, respectively (see Tables~\ref{table:cont} and \ref{table:CII})
\textit{Top row:} naturally weighted (0\farcs03, $\sim200$\,pc) \cont\ continuum maps. The contours are $-2\sigma$ (white contours) and $2\sigma, 3\sigma$ and $5\sigma$ (black contours). The maps show the continuum morphologies to be either compact and smooth (\Asixone, \Asixfive\ and \Aseventhree) or break up into apparent substructures on $\sim$ 200\,pc scales (\Afourseven). These 0\farcs03 resolution observations recover between 44 and 81\% of the continuum flux detected at lower resolution in ALMA Cycle 0/1 \citep{swinbank12,debreuck14,simpson15b}. 
\textit{Middle row:} intermediate resolution \cont\ continuum maps $uv$-tapered to 0\farcs05 ($\sim300$\,pc), showing the morphologies of the more extended emission in these sources. The contours are $-2\sigma$ (gray contours) and $2\sigma, 3\sigma, 5\sigma, 7\sigma$ and $9\sigma$ (black contours). The lower resolution images reveal more smooth structures.
\textit{Bottom row:} lowest resolution \cont\ continuum maps $uv$-tapered to 0\farcs2 ($\sim1.3$\,kpc) resolution, showing the most extended emission observable at this configuration. At this resolution the SMGs are unresolved and we recover between 80 and 100\% of the continuum emission.
}}
\label{fig:cont}
\end{figure*} 

\begin{table*}
\centering          
\begin{tabular}{l | c c c c | c c c c | c c}
\hline\hline
                  & \multicolumn{4}{c}{Natural weighting (0\farcs03)} & \multicolumn{4}{c}{Low-resolution (0\farcs2)} & \multicolumn{2}{c}{Sizes} \\
\hline       
Source                  & RMS                                                                & S/N & $S$  & Recovered & RMS                                                                & S/N & $S$ & Recovered &  FWHM$^{\text{uv}}$    & Aperture \\
                             & [$\mu$Jy\,beam$^{-1}$]  &        & [mJy] & flux           & [$\mu$Jy\,beam$^{-1}$]  &        & [mJy] & flux         &   [arcsec]  &     [arcsec]    \\ 
\hline                    
\Asixone               & 42 & 7.4 & $3.5\pm0.3$ & $81\pm12$\% & 0.32  & 7.7   & $6.5\pm0.2$   & $150\pm20$\% & $0.33\pm0.04$ & 0.40 \\ 
\Asixfive               & 42 & 7.5 & $3.0\pm0.2$ & $71\pm10$\% & 0.22  &  9.8  & $4.8\pm0.2$   & $110\pm10$\% & $0.30\pm0.04$ & 0.30 \\ 
\Aseventhree       & 27 & 8.3 & $2.9\pm0.2$ & $44\pm3$\%   & 0.16  & 11.7  & $5.4\pm0.2$   & $80\pm5$\% & $0.43\pm0.03$ & 0.36 \\ 
\Afourseven         & 47 & 6.6 & $5.6\pm0.3$ & $64\pm8$\%   & 0.25  &  20.8 & $10.2\pm0.2$ & $120\pm10$\% & $0.30\pm0.02$ & 0.30 \\
\hline
\end{tabular}
\caption{{\small Summary of the continuum properties of our sample at natural weighted (0\farcs03) and intermediate (0\farcs2) resolutions.
\textit{Column 5:} percentage of recovered flux from ALMA Cycle 0/1 data \citep{swinbank12,debreuck14,simpson15b}. 
\textit{Column 9:} percentage of recovered flux from ALMA Cycle 0/1 data \citep{swinbank12,debreuck14,simpson15b}. 
\textit{Column 10:} FWHM given by the Gaussian profile fit to the amplitude as a function of $uv$-distance}. 
\textit{Column 11:} optimized aperture size.}
\label{table:cont}
\end{table*}

In addition to the three ALESS sources we include \Afourseven\footnote{The numerical identifier for this SCUBA-2 source from \cite{simpson17} changed in the final version of the S2CLS UDS catalog \cite{geach17} with the source corresponding to UDS0051 in that work. This ALMA identified SMG is cataloged as AS2UDS0051.0 in Stach et al. in prep.} from the ALMA follow-up program of the SCUBA-2 Cosmology Legacy Survey \citep{geach17}.
A pilot study observed the 30 brightest SCUBA-2 sources in the $\sim1$\,deg$^{2}$ UKIDSS/UDS field \citep{simpson15b, simpson15a, simpson17}.
These 0\farcs3 resolution ALMA observations revealed a serendipitous detection of \CII\ emission at $350.78$\,GHz, establishing the redshift as $z=4.420$.
The now complete ALMA survey of $\gtrsim700$ sub-millimeter sources in the UDS field will be reported in Stach et al. (in prep.). 

\subsection{Physical properties}
\noindent
We determine the far-infrared luminosity of the galaxies in
 our sample by fitting modified black bodies to their spectral energy
 distributions (SEDs), including (deblended) 250, 350, 500\,$\mu$m
 flux densities (see \citealt{swinbank14}).  
 We adopt an average
 dust temperature of $T_{\text{d}}=50\pm4$\,K, a dust emissivity index
 of $\beta=1.5$ and assume the dust is optically thick at $\lambda=70\,\mu$m.  
 The choice of dust temperature is motivated by recent studies by \citealt{faisst17} and Cooke et al. (in prep.), both of which suggest that high-redshift galaxies with high specific star-formation rates have  higher characteristic dust temperatures than redshift $z\sim2$ SMGs \citep[$T_{\text{d}}\sim35$\,K e.g.,\ ][]{chapman05, swinbank12,weiss13}.
 Cooke et al. (in prep.) stack \textit{Herschel} PACs
 and SPIRE photometry (including from 100\,$\mu$m, 160\,$\mu$m,
 250\,$\mu$m, 350\,$\mu$m, 500\,$\mu$m) and ALMA $870\,\mu$m continuum
 measurements of thirteen $z\sim4.5$ ALMA SMGs with similar selection criteria to our sample and show that the ALMA SMGs at redshift $z\sim4.4$ have characteristic dust temperatures of $T_{\text{d}}=50\pm4$\,K.
We note that the far-infrared luminosity is sensitive to the dust temperature, where a lower dust temperature will result in a lower far-infrared luminosity. In Table~\ref{table:cont_cy0} we provide the far-infrared luminosities from from the best fit modified blackbodies.

\begin{table}
\centering          
\begin{tabular}{l c c c}
\hline\hline
Source name        & $M_{\text{dust}}$      & $M_{\text{gas}}$      & $M^{\text{[CII]}}_{\text{gas}}$\\
                             & [$10^8$\,M$_{\odot}$] & [$10^{10}$\,M$_{\odot}$] & [$10^{10}$\,M$_{\odot}$] \\
\hline                    
\Asixone               & $2.9\pm0.6$ & $2.6\pm0.9$ & $1.5\pm0.4$ \\ 
\Asixfive               & $2.8\pm0.6$ & $2.6\pm0.9$ & $3.2\pm0.8$\\
\Aseventhree       & $4.3\pm0.8$ & $3.9\pm1.3$ & $4.9\pm1.0$\\
\Afourseven         & $5.9\pm1.1$ & $5.3\pm1.8$& $2.6\pm0.7$\\
\hline
\end{tabular}
\caption{{\small  The estimated dust and gas masses based on the ALMA Cycle 0/1 observed \cont\ dust continuum extrapolated to rest-frame assuming $\beta=1.5$ and \CII\ fluxes. The dust masses (M$_{\text{dust}}$) are calculated using the \cont\ continuum flux, which are then scaled using a gas-to-dust mass ratio of    $90\pm25$ to achieve the gas masses (M$_{\text{gas}}$). Gas masses estimated using the \CII\ fluxes (M$^{\text{[CII]}}_{\text{gas}}$) are likewise listed, and agree   with the gas masses estimated using the dust mass.}}
\label{table:masses}
\end{table}

We calculate the dust masses using the measured continuum flux from 
ALMA listed in \ref{table:cont_cy0} and $M_{\text{d}}=S_{\nu}D_{\text{L}}^2/(\kappa
B_{\nu}(T_{\text{d}})(1+z))$, where $\kappa B_{\nu}(T_{\text{d}})$ is
the Planck-function modified by the dust absorption coefficient of
$0.076$\,m$^2$\,kg$^{-1}$ \citep{james02} which has been corrected from 
the rest-frame wavelength of $\sim160\,\mu$m to observed wavelength of $\sim870\,\mu$m assuming $\beta=1.5$, $D_{\text{L}}$ is the
luminosity distance and $S_{\nu}$ is the observed flux density at
frequency $\nu$ and adopt a characteristic dust temperatures of 50\,K and
 $\beta=1.5$ (Table~\ref{table:masses}).  Here we only use a single modified
blackbody, however, the dust mass for \Aseventhree\ was determined by
\cite{swinbank14} to be $9.3\pm0.6\times10^8$\,M$_{\odot}$ for a
multi-component model. This difference in masses is likely due
 to the different dust temperatures and assumed $\beta$-values,
 combined with the fact that the multi-component model trace a larger
 fraction of the dust mass at multiple temperatures. None of the
 other three SMGs have previously derived dust masses.

Adopting a single gas-to-dust mass ratio of
 $\delta_{\text{GDR}}=90\pm25$ \citep{swinbank14}, we estimate the
 gas masses (Table~\ref{table:masses}).  Also listed in
 Table~\ref{table:masses} are gas masses estimated using the \CII\ luminosities and the scaling relation:
 $M_{gas}=10\pm2\times(L_{\text{[CII]}}/L_{\odot})$
 \citep{swinbank12}. The two independent methods of estimating the
 gas masses result in masses agreeing within the uncertainties.

\section{Observations and reduction} \label{sec:obs}
\noindent
The four SMGs in our sample were observed with ALMA in Band 7 on 2015 November 9--14 for 22.7 to 40.7\,min on source, using 44 to 47 antenna in extended configurations, with the longest baselines being $\sim16.2$\,km (2015.1.00456.S).
The receivers were tuned such that one of the two spectral windows in the 7.5\,GHz side-band was centered to cover the expected frequency of the \CII\ emission line.
The FWHM of the ALMA primary beam is 18\arcsec at \cont.  
For the three ALESS sources, the QSOs J0522$-$3627, J0334$-$4008, and J0348$-$2749 were used as bandpass, flux and phase calibrator, while the QSOs J0238+1636 and J0217$-$0820 were used for \Afourseven. 

We used the Common Astronomy Software Application \cite[\textsc{casa},][]{mcmullin07} version 4.5.3 to reduce the data for all four sources and version 4.6.0 for further analysis. 
The ALMA pipeline was used for the reduction and delivered a high-quality product adopted for the analysis without further modifications. 
The data were imaged using the \textsc{clean} algorithm in \textsc{casa} with natural weighting (\textsc{ROBUST} $=2$) in order to recover as much of the extended emission as possible. 
We \textsc{CLEAN} the images to the RMS level of 27--47\,$\mu$Jy\,beam$^{-1}$ (see Table~\ref{table:cont}).
Using natural weighting and the full available $uv$-coverage results in a synthesized beam size of 0\farcs03$\times$0\farcs02 at position angle (P.A.) of 50--59$^{\circ}$ for all four SMGs.
We created maps at intermediate resolution by applying an outer $uv$-taper of 3500\,k$\lambda$ to the data when imaging.
This results in a synthesised beam size of 0\farcs05$\times$0\farcs04 at P.A. of 51--59$^{\circ}$ and an RMS of 35--60\,$\mu$Jy\,beam$^{-1}$.
Finally, we create our lowest resolution (0\farcs22$\times$0\farcs18) maps with RMS of 0.2--0.3\,mJy\,beam$^{-1}$ by applying an outer $uv$-taper of 350\,k$\lambda$ to the data when imaging.

We note that the observations were taken in ALMA's most extended configuration, which resulted in a well covered $uv$-plane for baselines $\gtrsim250$\,k$\lambda$, but poorer coverage at short baselines equivalent to the largest angular resolution (LAS) of $\sim0\farcs42$. 
This means that low surface-brightness emission, normally traced by shorter baselines (i.e.\ lower resolution) is difficult to detect. 
As we will show the dust continuum sizes of our sources are $\lesssim$LAS and so we recover most of the flux in the naturally weighted maps. 
However, the \CII\ emission in these sources appear more extended and so our high-resolution observations resolve out most of the emission.
This can result in an incomplete picture of the morphology and extent of the system and such high-resolution data therefore have to be interpreted with care. 

\begin{figure*}
\includegraphics[trim=1.1cm 0.4cm 0.2cm 0cm, clip=true,scale=0.6,angle=90]{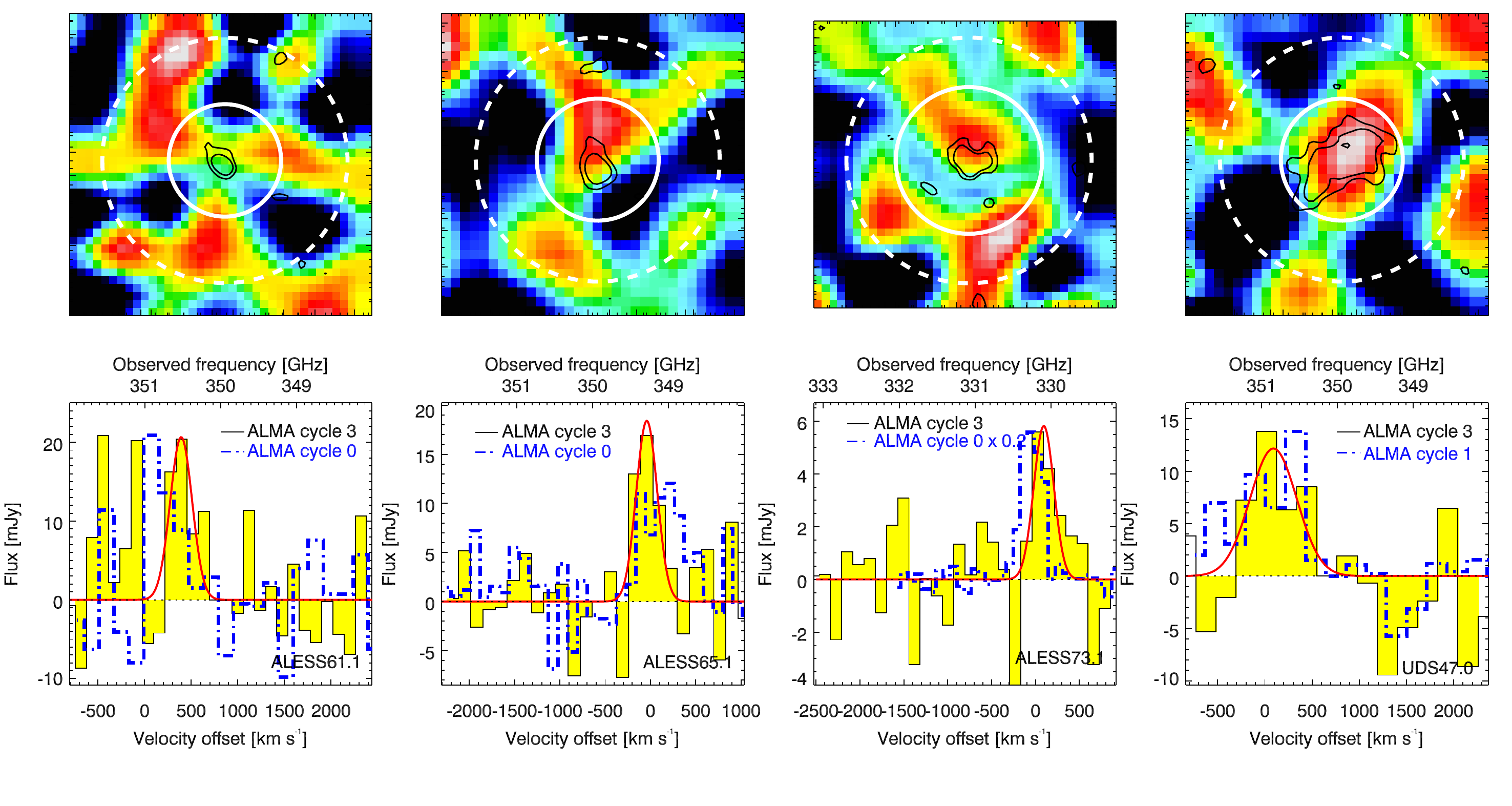}
\caption{{\small \CII\ moment zero maps and comparison of the continuum-subtracted \CII\ emission-line spectra from the new 0\farcs03 resolution data and the lower resolution (see Table~\ref{table:cont_cy0}) ALMA Cycle 0/1 observations. The moment-zero \CII\ are obtained using a $uv$-taper of 500\,k$\lambda$. Overlaid in white solid and white dashed circles are the $uv$-derived sizes estimated for the \cont\ dust continuum emission and the \CII\ emitting gas, respectively (see Tables~\ref{table:cont} and \ref{table:CII})
The spectra from the 0\farcs03 resolution observations have been binned up to $130-210$\,km\,s$^{-1}$ per channel. 
Only by using large optimized apertures (compared to the resolution), do we recover the \CII\ flux seen in the shallower Cycle 0/1 observations with large uncertainties. This indicates that the \CII\ emission is very extended in these sources and relatively smooth and hence our 0\farcs03 resolution observations are resolving out the bulk of the extended \CII\ flux emission in these sources. 
The ALMA Cycle 0 spectrum for \Aseventhree\ from \cite{debreuck14} has been scaled down by a factor of five.
The line peak on our Cycle 3 observations of \Asixone\ is shifted by $\sim400$\,km\,s$^{-1}$ and is dominated by high noise spikes on the blue side of the line.
The shift in line center seen in the other sources may be due to low signal-to-noise or the fact that the small scale structure detected in our high-resolution observations is not uniformly distributed within the sources.}}
\label{fig:compare_spec}
\end{figure*}

\section{Analysis} \label{sec:analysis}
\noindent
The resolution of the naturally weighted maps (i.e.,\ $0\farcs03$) enables us to search for sub-structures in these sources on $\sim200$\,pc scales. 
Figure~\ref{fig:cont} shows the \cont\ continuum maps at this resolution, which reveal a broad range of apparent morphologies: smooth and compact sources in \Asixone, \Asixfive\ and \Aseventhree, and extended structure that appears to break up into $\sim200$\,pc scale clumps in \Afourseven. 

The peaks of the \cont\ continuum for the four sources are detected at $7.4-8.3\sigma$ and with an optimized aperture size determined by using a curve of growth we recover between $44\pm3$ and $81\pm12$\% of the integrated flux density measured from the low resolution ALMA Cycle 0/1 observations \citep{swinbank12,debreuck14,simpson15b}. 
This suggests that our maps are missing a fraction of the emission from the most extended low surface brightness dust on scales $\gtrsim200$\,kpc.

To try to recover more of the extended emission in our maps we also applied an outer $uv$-taper to the data, thereby giving a greater weight to the shortest baselines at the cost of reduced resolution. 
The middle row of Fig.~\ref{fig:cont} shows the intermediate resolution continuum maps with an outer $uv$-taper of 3500\,k$\lambda$ at a resolution of $\sim300$\,pc. 
These maps show more of the extended lower surface brightness emission, and have a higher signal-to-noise ratio than the naturally weighted maps. 
We therefore fit exponential profiles to these maps and find mean \sersic\ indices of $n=1.1\pm 0.1$, which is in agreement with the indices from $\sim0\farcs15$ resolution imaging of 16 ALESS SMGs \citep{hodge16}. 

To maximise the recovered flux we applied an outer $uv$-taper of 350\,k$\lambda$, resulting in our lowest resolution maps of $\sim1.3$\,kpc (see bottom row of Fig.~\ref{fig:cont}), which recover 80 to 100\% of the continuum flux detected in the ALMA Cycle 0/1 observations \citep[][see Table~\ref{table:cont}]{swinbank12, debreuck14, simpson15b}.
It is only for \Aseventhree\ that there is an indication that we are still resolving out some flux in these low resolution maps, where we recover $80\pm5$\% of the flux detected in ALMA Cycle 0/1. 
In later calculations we treat the flux recovered at our low resolution data as the total flux. 

\subsection{\CII\ emission lines}
\begin{table*}
\centering          
\begin{tabular}{l c c c c c c c}
\hline\hline       
Source                  &  RMS & S/N & $SdV_{\text{[CII]}}$ & FWHM$^{\text{line}}_{\text{[CII]}}$ & FWHM$^{\text{uv}}_{\text{[CII]}}$ & Aperture & Recovered\\
                             &  [mJy] &        &  [Jy\,km\,s$^{-1}$]   &              [km\,s$^{-1}$]                     &            [arcsec]             &   [arcsec]      &      [\%]       \\ 
\hline                    
\Asixone               &  7.4     & 2.7  & $4.6\pm1.7$         &    $280\pm110$  & $1.1\pm0.4$    &  0.72     & $180\pm80$                 \\ 
\Asixfive               &  4.0     & 4.7  &   $4.9\pm1.0$        &     $270\pm70$   & $0.6\pm0.2$    &  0.6       & $90\pm20$                   \\ 
\Aseventhree       &  2.0     & 3.0 &   $1.6\pm0.5$        &     $270\pm110$  & $0.7\pm0.1$     &  0.72    & $22\pm7$                     \\ 
\Afourseven         &  4.4     & 3.1  &   $6.8\pm1.8$        &     $590\pm250$  & $0.3\pm0.1$     &  0.6      & $160\pm50$                \\ 
\hline
\end{tabular}
\caption{{\small Properties of the \CII\ emission lines detected in the 0\farcs03 resolution observations,
\textit{Column 2:} RMS of the \CII\ spectra.
\textit{Column 3:} Signal-to-noise ratio of the \CII\ emission lines.
\textit{Column 4:} Velocity integrated line fluxes. 
\textit{Column 5:} FWHM of the \CII\ line velocity width.
\textit{Column 6:} Spatial FWHM given by the Gaussian fit to the amplitude as a function of $uv$-distance.
\textit{Column 7:} Diameter of the aperture used to measure the line flux.
\textit{Column 8:} percentage of recovered line flux.}}
\label{table:CII}
\end{table*}

\noindent 
Lower resolution ALMA studies have demonstrated that these four SMGs are all bright \CII\ emitters \citep{swinbank12,debreuck14}. 
By using $uv$-tapering we recovered between 80-100\% of the continuum flux detected in ALMA cycle 0/1. 
However, $uv$-tapering only recovers emission in the image-plane, and does not improve the signal-to-noise of the spectral line. 
We therefore search for \CII\ emission in our 0\farcs03 observations and select extraction apertures to maximize the recovered signal-to-noise of the line emission. This results in the recovery of modest significant ($2.7-4.7\sigma$) \CII\ emission lines (see Fig.~\ref{fig:compare_spec}), with measured RMS values for the spectra of $2.0-7.4$\,mJy in 130--210\,km\,s$^{-1}$ channels (see Table~\ref{table:CII}). 
Figure~\ref{fig:compare_spec} shows the moment-zero maps with a $uv$-taper of 500\,k$\lambda$ ($0\farcs17\times0\farcs16$) and compares the recovered \CII\ emission with the spectra from the lower resolution observations from ALMA Cycle 0 and 1 \citep{swinbank12,debreuck14}. 

\Asixone\ and \Asixfive\ were detected in \CII\ emission in ALMA Cycle 0 and we recover between 90 and 100\% of the velocity integrated line flux at 0\farcs03 resolution, using apertures with diameters of 0\farcs6 and 0\farcs72 respectively. 

For \Aseventhree\ at 0\farcs03 resolution (Fig.~\ref{fig:compare_spec}) our observations recover only $\sim20\%$ of the peak flux emission in the 0\farcs5 resolution map from \cite{debreuck14}. 
To compare with the emission line profiles detected in the 0\farcs5 resolution data \citep{debreuck14}, we simply scale the peak of the 0\farcs5 resolution \CII\ line to that of the 0\farcs03 emission line (i.e.,\ multiply by 0.2). 
This results in the red wing of the \CII\ emission line we detect being consistent with the red wing of the \CII\ emission line from the 0\farcs5 resolution data before down-scaling.

Emission from \CII\ was detected for \Afourseven\ as a very broad line at $\sim351$\,GHz in the shallower 0\farcs3 data from ALMA Cycle 1 \citep{simpson15a,simpson17}. 
In our deeper 0\farcs03 resolution observations we detect a broad $\sim4\sigma$ \CII\ emission line. 
Adopting an optimized aperture size of 0\farcs6 we recover the full flux seen in the shallower low resolution observations from ALMA Cycle 1. 

\subsection{Size estimates from $uv$-plane fits}\label{sec:uv-fit}
\begin{figure*}
\centering
\includegraphics[trim=8.1cm 0.1cm 0.9cm 1.1cm, clip=true,scale=0.76,angle=90]{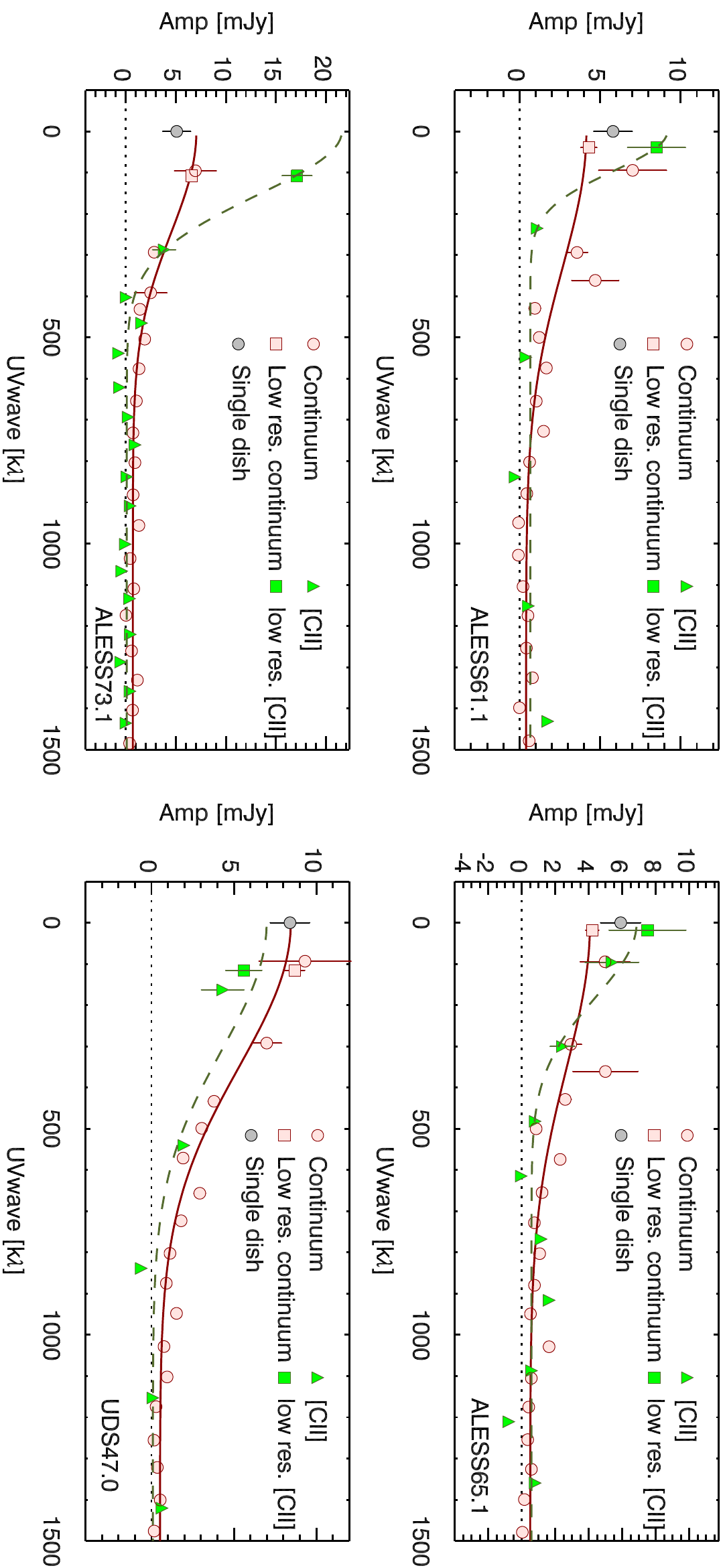}
\caption{{\small 
Visibility amplitudes as a function of the $uv$-distance for all four SMGs in the sample for the \cont\ dust continuum and \CII\ emission. 
The amplitudes for the continuum are extracted by radially averaging the visibilities in 75\,k$\lambda$ bins over the full frequency range. 
For the \CII\ emission in \Asixone\ and \Afourseven\ a larger binning of 300\,k$\lambda$ had to be applied for the radially averaging of the visibilities, while a binning of 75\,k$\lambda$ was applied to \Asixfive\ and \Aseventhree. 
The line visibilities here cover the observed spectral range of the \CII\ emission of 350--352\,GHz for \Asixone\ and \Afourseven, 348--350\,GHz for \Asixfive, and 329--311\,GHz for \Aseventhree. 
The $uv$-coverage is better sampled at long baselines than on shorter ones for our ALMA configuration. 
The half-Gaussian fits to the continuum emission are overlaid in red and the half-Gaussian fits to the \CII\ emission are shown in green. 
We also plot the fluxes determined from the lower resolution data from ALMA Cycle 0 and 1. 
The \cont\ dust continuum and \CII\ emission of the SMGs are resolved in our observations and the \cont\ dust continuum sizes of the sources are listed in Table~\ref{table:cont} and \ref{table:CII}.
Despite the sparse $uv$-coverage at $<250$\,k$\lambda$ it is evident that the extent of \CII\ emission is greater than or equal to the extent of the \cont\ dust continuum emission in all four SMGs.
}}
\label{fig:amp_uv}
\end{figure*}

\subsubsection{Continuum}
\noindent
The natural-weighted resolution \cont\ dust continuum maps recover between 44 to 81\% of the continuum flux detected at lower resolution. 
This suggests that around $40\pm20$\% of the flux has been resolved out at 0\farcs03 resolution compared to the Cycle 0/1 maps. 
To estimate the size of the rest frame 160\,$\mu$m emission in the SMGs we determine the behavior of the amplitude as a function of $uv$-distance.
We do this by first aligning the phase center of our cubes with the source position listed in Table~\ref{table:cont_cy0} and then radially average the data in 75\,k$\lambda$ bins to extract the amplitudes as a function of $uv$-separation. 
For the continuum a binning of 75\,k$\lambda$ is the most optimal to minimize the scatter, however, the overall trend of the amplitude as a function of $uv$-distance is independent of the binning. 
Figure~\ref{fig:amp_uv} shows the amplitude as a function of the $uv$-distance for the inner 1500\,k$\lambda$ in each of the four SMGs.

For a point source the observed amplitudes will be constant as a function of $uv$-distance, while for an extended source the amplitude declines at larger $uv$-distances. 
Hence the physical size of the source can be determined from the FWHM of a half-Gaussian profile fit to these $uv$-profiles. 
In that case the total flux is represented by the peak value of the half-Gaussian fit.  
As Fig.~\ref{fig:amp_uv} shows, the amplitude declines as a function of $uv$-distance for the continuum emission in all four SMGs and they are therefore consistent with a centrally peaked brightness profile, meaning that the sources are resolved.  
We add the low resolution observations from Cycle 0/1 to the plots at the $uv$-distance corresponding to the LAS of the observations and the single dish flux at 0\,k$\lambda$ as these represent our best estimate of the total integrated flux. We fit a half-Gaussian profile plus a constant (representing a point source) to the amplitudes in order to measure the physical size of the emission, and to establish whether a point source is present.
The fitted FWHMs converted into physical sizes are listed in Table~\ref{table:cont}. 

We find that the continuum point source components of the fits are non-zero for all four sources, with flux densities of $0.4-0.7$\,mJy. 
This suggests that on average about $\sim14$\% of the total continuum flux in each source is emitted from a component with a size $\lesssim200$\,pc. 

There is a published size for the 330\,GHz continuum reported of  \Aseventhree\ $0\farcs29\pm0\farcs06$ \citep{debreuck14}.
To compare with this we derive a size from a single Gaussian fit without a point source of $0\farcs38\pm0\farcs05$ which is consistent.

As already stated we only recover the total continuum flux density from the low-resolution observations in the $uv$-tapered map by applying an outer taper of $\sim350$\,k$\lambda$.
Figure~\ref{fig:amp_uv} illustrates that, since the amplitudes only diverge from the constant value of $0.4-0.7$\,mJy at $uv$-distances of $\lesssim350$\,k$\lambda$, only a strong $uv$-taper gives enough weight to the shortest baselines to lower the resolution sufficiently to make a significant difference in the recovered flux density. 

\subsubsection{\CII\ emission}
\noindent
We recover \CII\ emission lines in all four SMGs, but only at low significance (2.7--4.7$\sigma$).
To determine whether the \CII\ emission is resolved and to estimate its extent, we extract the amplitude as a function of the $uv$-distance for the spectral channels spanning the \CII\ emission. 
We align the phase centre to the same position as for the continuum and fit a zero-order polynomial in the $uv$-plane to the line free channels to determine the continuum level. 
We then subtract the fitted continuum in the $uv$-plane and extract the amplitude as a function of $uv$-distance for the spectral channels spanning the \CII\ emission. 
Estimates of the physical extent of the \CII\ emission use the same method as for the continuum, i.e.,\ by fitting a half-Gaussian profile plus a point source component to the amplitude as a function of the $uv$-distance, and converting the fitted FWHMs of the Gaussian profile to arc-seconds. 
Figure~\ref{fig:amp_uv} shows the profiles of the continuum emission and the \CII\ emission, with the FWHMs given in Table~\ref{table:CII}. 
Note that besides the poor sampling of the inner part of the $uv$-plane, the visibilities for the \CII\ data are derived from only a limited spectral range and so contain fewer data points, and we therefore have to apply a larger binning of 300\,k$\lambda$ for \Asixone\ and \Afourseven\ and 75\,k$\lambda$ for \Asixfive\ and \Aseventhree. 
\cite{debreuck14} measure the \CII\ emission to have a FWHM of $\sim0\farcs64$ in \Aseventhree, which is consistent with our measurements (FWHM$=0\farcs7\pm0\farcs1$).

Our data are not well enough sampled to establish whether unresolved \CII\ emitting components are present in these four SMGs. 
The lower sampling of visibility points in the \CII\ data also means that the measured sizes are more uncertain than that measured for the \cont\ continuum. 
We include the lower resolution observations from Cycle 0/1 in our fits yielding estimated FWHMs of $0\farcs3 - 1\farcs1$, comparable or larger than the LAS of $\sim0\farcs4$ recoverable at the antenna configuration of our Cycle 3 observations.

\section{Discussion} \label{sec:disc}
\subsection{Continuum and \CII\ sizes}\label{sec:sizes}
\noindent
From the half-Gaussian profile fits in Fig.~\ref{fig:amp_uv} we measure the median continuum size to be 0\farcs32$\pm$0\farcs03 and the \CII\ size to be $0\farcs65\pm0\farcs15$ (see Tables~\ref{table:cont} and \ref{table:CII}). 
The measured size ratio for our sample suggest that on average the \CII\ emitting gas is $2.1\pm0.4$ times more extended than the continuum emitting dust. 
The fact that the continuum sizes are smaller than the LAS, allows us to recover between 80 to 100\% of the flux detected in ALMA cycle 0/1. 
The \CII\ sizes, however, are larger than the LAS, meaning it is not possible to recover the emission distributed on scales larger than the LAS. This therefore results in  low signal-to-noise  \CII\ emission lines and low significance moment-zero maps (Fig.~\ref{fig:compare_spec}).

\begin{figure}
\centering
\includegraphics[trim=15.8cm 2.8cm 2.9cm 10cm, clip=true,scale=0.5,angle=90]{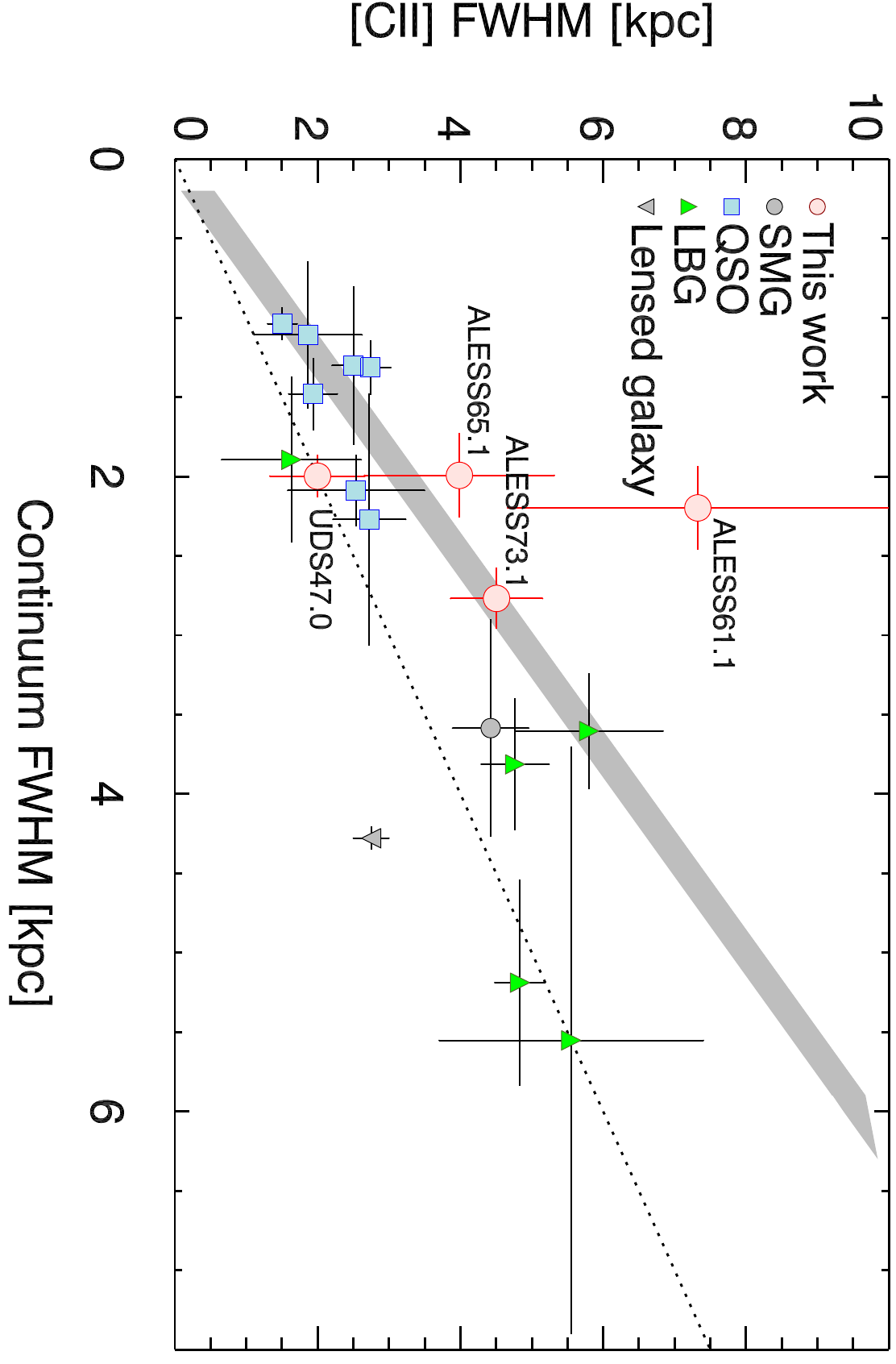}
\caption{{\small Spatial FWHM of the rest frame 160\,$\mu$m continuum emission versus the FWHM of the \CII\ emission for the four SMGs in this work and high-redshift ($3.1<z<7.1$) galaxies in the literature with similar measurements \citep{wang13,capak15,kimball15,nesvadba16,umehata17, diaz-santos16,venemans17,jones17}. We see that the \CII\ emitting gas is more extended than the rest frame 160\,$\mu$m emitting dust for the majority of the galaxies. 
The dotted line is the 1:1 relation,} while the gray shaded region shows the weighted mean of the \CII\ to continuum size of $1.6\pm0.4$.}
\label{fig:size}
\end{figure}

Figure~\ref{fig:size} compares our estimated \CII\ and rest-frame 160\,$\mu$m continuum sizes for our SMGs. 
It also shows the \CII\ and rest-frame 160\,$\mu$m dust continuum sizes for four quasars at $z=4.6-7.1$ \citep{wang13, kimball15, diaz-santos16, venemans17}, a starburst galaxy at $z=3.4$ \citep{nesvadba16}, a Lyman-$\alpha$ Blob at $z=3.1$ \citep{umehata17} and LBGs at $z=5.3-6.1$ \citep{capak15,jones17}. 
The \CII\ and rest-frame 160\,$\mu$m dust continuum observations have been taken at the same spatial resolution in each source, but this varies between 0\farcs2 and 1\arcsec. 
These observations appear to support the conclusion that \CII\ emitting components are more extended than the rest-frame 160\,$\mu$m dust components in a majority of the systems. 

Although the resolution of these studies is $\sim6-30$ times lower than our observations, the relative sizes of the \CII\ and rest-frame 160\,$\mu$m dust emission still suggest that the \CII\ emitting gas is more extended than the rest-frame 160\,$\mu$m continuum. 
The weighted mean of the \CII\ to rest-frame 160\,$\mu$m dust continuum size, including our four SMGs and the comparison sample, is $1.6\pm0.4$.
Only three of the eighteen galaxies have apparently larger rest-frame 160\,$\mu$m continuum than \CII\ sizes and therefore lie off this relation, and only one of these is significantly different: a lensed starburst galaxy, where the relative sizes are sensitive to the details of the lens model. 
The fact that the majority of the galaxies follow a trend, although they are very different populations with different gas masses and AGN luminosities, suggests that these global physical parameters are unlikely to account for the observed size differences. 
This means that the nature of the dominant heating source (whether, for example, it is AGN or starburst activity) does not appear to significantly influence the relative size of the rest-frame 160\,$\mu$m dust continuum and \CII\ emitting gas.

At these high redshifts ($z\simeq4.5$) the temperature of the cosmic microwave background (CMB) is $\simeq15$\,K. 
This means that if the star-forming dust has similar temperature to the CMB, it will not be detectable \citep{dacunha13,zhang16}. 
We note that given that the dust temperature is higher than the background CMB, this means that the CMB is unlikely to be the reason why the \CII\ emission is $1.6$ times more extended than the rest-frame 160\,$\mu$m dust emission. 

\subsection{Velocity gradients and dynamical masses}\label{sec:dyn} 
\noindent
The \CII\ emission line is one of the brightest cooling lines of the interstellar medium and traces the ionized, neutral and molecular gas.
It is therefore a good tracer of the gas dynamics in high-redshift galaxies \citep[e.g.,][]{carniani13,capak15}. 
Only \Aseventhree\ and \Afourseven\ have low-resolution observations from Cycle 0/1 deep enough to allow us to search for possible velocity gradients.
From the study of \cite{debreuck14} it is already known that the gas in \Aseventhree\ has a rotating configuration, and the broad line of \Afourseven\ suggests that a velocity gradient may also be present there. 

To investigate the velocity gradient in these two SMGs we make moment-zero maps (i.e.,\ narrow-band images) in the low resolution (from ALMA Cycle 0/1, see Table~\ref{table:cont_cy0}) continuum-subtracted cube of the channels covering the \CII\ emission. 
We make two independent maps; one of the redshifted half of the line and the other of the blueshifted half. 
These cover $\pm400$\,km\,s$^{-1}$ for \Afourseven\ and $\pm200$\,km\,s$^{-1}$ for \Aseventhree.
We find that the peak of the \CII\ emission shifts by $0\farcs25\pm0\farcs04$ ($\sim1.7\pm0.3$\,kpc) between the red and blue halves of the line for \Afourseven\ and $0\farcs24\pm0\farcs01$ ($\sim1.5\pm06$\,kpc) for \Aseventhree.
This implies a velocity gradient across the \CII\ emitting gas in both SMGs.

Having established the presence of a velocity gradient and using a disk model for the dynamics, we estimate the dynamical masses ($M_{\text{dyn}}\sin(i) = R\times v^2/G$) of \Aseventhree\ and \Afourseven\ within a region given by twice the size of the \CII\ sizes listed in Table~\ref{table:cont} corresponding to $R=5$\,kpc  and $R=4$\,kpc for \Aseventhree\ and \Afourseven, respectively.  
Using the line widths of the \CII\ lines detected in ALMA Cycle 0/1 (see Table~\ref{table:cont_cy0}), this yields dynamical masses of $3.7\pm0.7\times10^{10}\sin(i)$\,M$_{\odot}$ for \Aseventhree\ and $20\pm4\times10^{10}\sin(i)$\,M$_{\odot}$ for \Afourseven. 
Using a similar disk model \cite{debreuck14} estimate an inclination angle of $i=50^{\circ}\pm8$ for \Aseventhree, which is similar to the average inclination angle calculated by \cite{law09}.
By assuming the same inclination angle for \Afourseven\ and the gas masses listed in Table~\ref{table:masses}, we estimate an average gas mass fraction within the half mass radii assumed to calculate the dynamical masses of $0.4\pm0.2$.  
This is in agreement with the result from \cite{tacconi17} for redshift $\lesssim4$ star forming galaxies when converting to the same units.

\subsection{Morphologies}\label{sec:morph}
\begin{figure*}
\centering
\includegraphics[trim=8.2cm 0.7cm 7.2cm 0.7cm, clip=true,scale=0.63,angle=90]{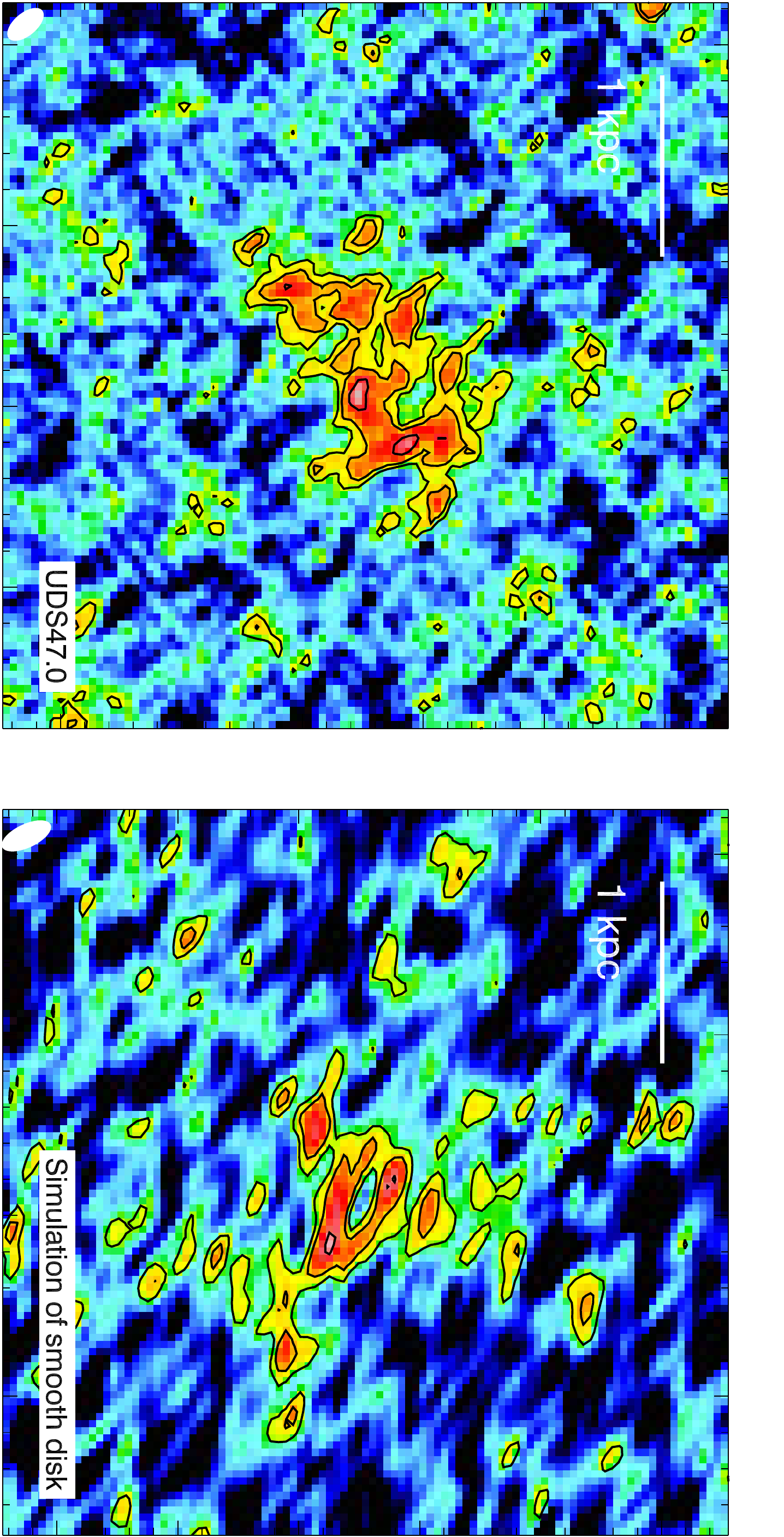}
\caption{{\small \textit{Left:} The 0\farcs03 resolution \cont\ continuum map of \Afourseven\ showing apparent clumps with sizes comparable to the synthesised beam ($\sim200$\,pc). 
\textit{Right:} An example of a simulated \cont\ continuum map of a smooth exponential disk observed with the same observational parameters as the data shown in the map to the left.
Both maps show similar apparent structures, but the right-hand simulated map is known to be a smooth exponential disk.
Analysis of the flux distribution of the pixels reveal that there is a 70\% probability that the map of \Afourseven\ is drawn from the simulated library of models. It is therefore not possible to rule out the hypothesis that \Afourseven\ is a smooth disk from our observations, and so we suggest that the apparent structures identified by eye may be misleading.
}}
\label{fig:sim}
\end{figure*}

\noindent
Figure~\ref{fig:cont} demonstrates that three SMGs in our sample (\Asixone, \Asixfive, and \Aseventhree) show smooth \cont\ dust continuum morphology, while one (\Afourseven) appears to have structure. 
However, the apparent structures seen for \Afourseven\ have significance levels of just $3.9-5.4\sigma$.
Similarly, a recent high-resolution (0\farcs03) resolution \cont\ dust continuum study by \cite{iono16} of three $z\sim4.3$ AzTEC SMGs claimed to reveal about $40$ $\geq3\sigma$ 200\,pc clumps. 
These visually identified structures are similar to the features we see in \Afourseven. 
 
To test whether the structures in \Afourseven\ are likely to be real, or if they could arise from noise in smooth disk light profiles we model a set of observations of smooth profiles. We use the \textsc{casa} tasks \textsc{simobservations} and \textsc{simanalysis} to create a library of simulated interferometry observations of exponential disks as they would appear if observed with ALMA in the same configuration as used for our observations and with similar noise properties (following the example of \citealt{hodge16}).
Our 50 input models of smooth exponential disk models have \sersic\ indices of $n=1$ (comparable to what we see in our sample) and flux densities and sizes of $8.7\pm0.6$\,mJy and $0\farcs28\pm0\farcs03$, as seen for \Afourseven\ \citep{simpson15b}.
The results of these simulations also reveal apparently clump-like structures (Fig.~\ref{fig:sim}).
While these structures qualitatively look similar to those seen in \Afourseven, we attempt to quantitatively compare the flux distribution between the simulated maps and the observed map for \Afourseven. 
We do this by fitting single smooth profiles (with the \sersic\ index as a free parameter, to the simulated maps), and subtracting the best fit model. 
For the central part of the residual image, the number of pixels as a function of the flux has a Gaussian profile with a tail of excess emission at positive values. 
This excess flux should represent the emission seen in possible structures, and we isolate it by subtracting a one-dimensional-Gaussian profile fit to the histogram.

We apply this analysis to both the simulated smooth disks, and the observation of \Afourseven. 
A Kolmogorov-Smirnov test comparing the average of the residual pixel distribution from the simulated smooth disks and that of the observations, reveals that the likelihood of the observed map being drawn from the simulated library of smooth disks is $\sim70\%$. 
Applying the same analysis for the three other SMGs in our sample, show that they are consistent with smooth morphologies. 

Our simulated library of smooth disks in combination with the apparently smooth morphology of three out of four SMGs in our sample, illustrates that smooth disks can appear to have substructures when observed at high-resolution and with sparse coverage of the inner part of the $uv$-plane.
We conclude that it is therefore not possible to rule out the hypothesis that all four SMGs in our sample are smooth exponential disks. 

We note that the structures identified by \cite{iono16} in their sources have similar significances to those seen in \Afourseven. 
Moreover only $\sim30$\% of the continuum flux detected at 0\farcs7 resolution with the SMA \citep{iono16,younger08} has been recovered in their 0\farcs03 ALMA maps, with less than $20$\% of that in the claimed clumps.
The fact that the resolution and depth of the observations in \cite{iono16} are similar to our 0\farcs03 maps and that their claimed structures contain only a small fraction of the total flux, casts doubt on whether their clumps are real structures either. 
  
\subsection{Expected size of clumps}\label{sec:clumpsize}
\noindent
As noted above, one of the sources in our sample (\Afourseven) appears to have a clumpy morphology in our high resolution ALMA continuum maps, however we have suggested that this is not statistically significant. 
Nevertheless, we can also ask if we should expect to see sub-structures at this resolution, given the estimated star-formation rate surface densities of these galaxies.

The average sizes of star-forming clumps in a self-gravitating gas disk are given by the Jeans length,
\begin{align}
\lambda_J\approx\frac{\sigma^2}{G\Sigma_{\text{gas}}},
\end{align}
where $G$ is the gravitational constant, $\Sigma_{\text{gas}}$ is the gas surface density, and $\sigma$ the velocity dispersion of the gas within a clump \citep{toomre64}.
We estimate $\Sigma_{\text{gas}}$ using:
\begin{align}
\left( \frac{\Sigma_{\text{SFR}}}{\text{M}_{\odot}\,\text{yr}^{-1}\,\text{kpc}^{-2}}\right)= A {\left(\frac{\Sigma_{\text{gas}}}{\text{M}_{\odot}\,\text{pc}^{-2}}\right)}^n,
\end{align}
where $A=1.5\times10^{-4}$ and $n\simeq1.5$ \citep{kennicutt98,swinbank12}.
For clumps to have a size of $\gtrsim200$\,pc (and thus be observable at the resolution of our observations at the estimated gas surface density), the velocity dispersions within the gas disk have to be $\sigma\gtrsim60$\,km\,s$^{-1}$ for \Asixone, which has the lowest estimated gas surface density of our sample, and $\sigma\gtrsim85$\,km\,s$^{-1}$ for \Afourseven, which has the highest  gas surface density.

A recent high resolution ($\sim0\farcs03$) observation of the lensed SMG SDP.81 \citep{almapartnership15,hatsukade15} measured the velocity dispersions in regions within the gas disk in this system to be in the range $11-35$\,km\,s$^{-1}$ \citep{swinbank15}. 
The velocity dispersion of the gas disk in \Aseventhree\ was likewise estimated to be $40\pm10$\,km\,s$^{-1}$ \citep{debreuck14}. 
Thus the required velocity dispersions to observe clumps at 200\,pc resolution are 1.5--2 times higher than that observed in other SMGs. 
Hence if our sources have velocity dispersions comparable to that observed for other SMGs, then any clumps in their gas disks would have sizes below the resolution limit of our ALMA observations. This suggests that the clumps in \Afourseven, if real, are unlikely to represent self-gravitating physical structures. 
  
\subsection{The \CII\ deficit}
\begin{figure*}
\centering
\includegraphics[trim=15.8cm 2.95cm 2.9cm 0.6cm, clip=true,scale=0.68,angle=90]{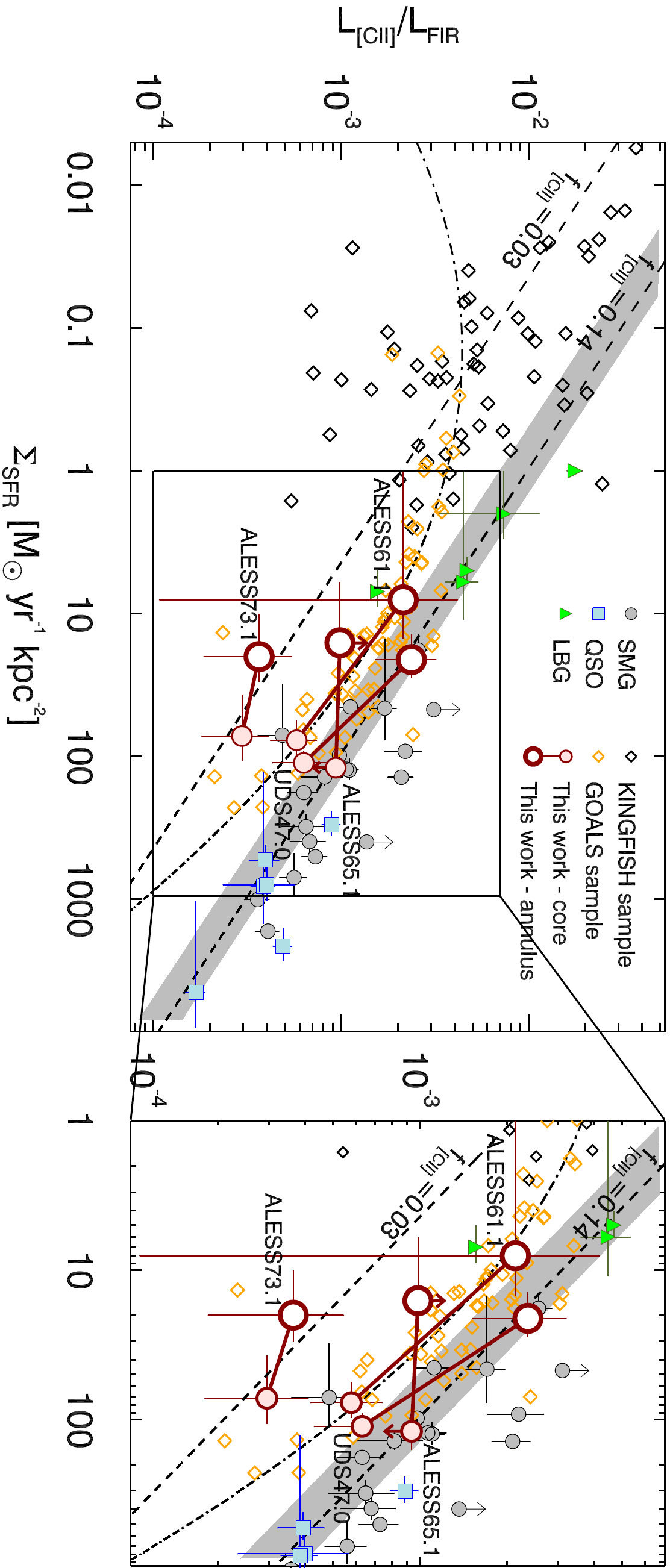}
\caption{{\small The \LCII\ to \LFIR\ ratio as a function of the star-formation rate density for local galaxies in the KINGFISH and GOALS \citep{armus09, diaz-santos17,lutz16}, a sample of high-redshift galaxies \citep{wang13, kimball15, capak15,gullberg15, diaz-santos17,nesvadba16,umehata17,spilker16, jones17, venemans17}, and our sample of four $z\sim4.5$ SMGs. 
We note that the sizes used to achieve the star-formation rate densities for the GOALS sample are the 70$\mu$m effective radii under the assumption of a uniform dust temperature.
The inner parts of our four galaxies are plotted as red-circles and the outer annuli as stars. The respective core and annuli measurements are connected by solid lines. 
For three SMGs in the sample the \densfr\ is lower in the outer annulus and the \LCII/\LFIR\ ratios is higher than for the core. 
This suggests that the \CII\ deficit is a local process. 
The \LCII/\LFIR\ ratio as a function of the FIR surface density and the mass fraction of \CII\ compared to the total mass ($f_{\text{[CII]}}$) from \cite{munoz16} is plotted as the gray shaded area, assuming $f_{\text{[CII]}}=0.10-0.17$, and the curved power law fitted to galaxies in GOALS from \citep{diaz-santos17} is plotted as the dot-dashed curve.
We note that both the high and low redshift samples exhibit a large scatter around the two models. 
Our data agrees with both models and so it is not possible to distinguish which of the two models is closer to the true explanation.
}}
\label{fig:sfrdens}
\end{figure*}

\noindent
We now turn to the overall energetics of these systems and their cooling. 
As noted earlier, emission from \CII\ is a major contributor to the gas cooling, carrying 0.1--1\% of the far-infrared luminosity in luminous starburst galaxies \citep{stacey91, brauher08, gracia_carpio11}. 
The \CII\ to far-infrared luminosity ratio (\LCII/\LFIR) varies with far-infrared luminosity in local galaxies, such that the most far-infrared-luminous galaxies have a lower \LCII/\LFIR\ ratio.
Figure~\ref{fig:sfrdens}  shows \LCII/\LFIR\ as function of star-formation rate surface density for the low-redshift KINGFISH sample \citep{kennicutt11, dale16,croxall17}, the GOALS sample \citep{diaz-santos13,lutz16}, and a high-redshift sample of SMGs \citep[][typically lensed]{gullberg15,spilker16}.

 At the highest star-formation rate surface densities (i.e.,\ typically smaller sizes, FWHM $\sim1-3$\,kpc) we see that the lowest \LCII/\LFIR\ ratios, are frequently associated with active galactic nuclei (AGN), while normal star-forming galaxies (including Lyman Break galaxies) have higher \LCII/\LFIR\ ratios.

We see in Fig.~\ref{fig:sfrdens} that the \LCII/\LFIR\ ratios for our sample agree with the high redshift comparison sample, but both show a large scatter when compared to the local galaxies from KINGFISH and GOALS. 
Figure~\ref{fig:sfrdens} shows that the high redshift sources in general have higher \LCII/\LFIR\ compared to local galaxies at a fixed far-infrared luminosity. 

To investigate if the \CII\ deficit is due to a local or global process, we plot two points for each of our SMGs: a core measurement from an aperture the same size as the \cont\ dust continuum (Table~\ref{table:cont}); and an annulus the size of the optimised \CII\ aperture listed in Table~\ref{table:CII}. 
The \CII\ luminosities are calculated by extracting two spectra, one within the continuum aperture and one within the \CII\ aperture; the \CII\ luminosity in the annulus is then given by the difference between the two luminosities. 
Note that for \Asixfive\ the \CII\ emission line is undetected in the core, meaning that the \LCII/\LFIR\ ratio for the core is an upper limit and therefore a lower limit within the annulus.
We scale the far-infrared luminosities and star-formation rates according to the fraction of emission we recover within the \cont\ dust continuum apertures (see Table~\ref{table:cont}), and assume that the remaining fraction originates from the annulus. 
The star-formation rate surface densities are then calculated using the areas of the continuum apertures for the core measurement and the difference between the \CII\ and continuum apertures for the annuli.

We use these measurements to investigate the variations of the \LCII/\LFIR\ ratio within our SMGs.
The expanded part of Fig.~\ref{fig:sfrdens} shows the trend between the core measurements and the annuli.
All four of the galaxies show the same behavior: the core has lower \LCII/\LFIR\ and higher star-formation rate surface density than the surrounding annulus.
This follows the relation seen by \cite{smith17} for local star-forming galaxies. 
A higher star-formation rate surface density in the core regions compared with the outer annuli is expected, however, the same expectation is not true for the \LCII/\LFIR\ ratio. 

The fact that our SMGs follow the same trend seen by for example \cite{smith17}, implies that the \CII\ deficit in SMGs is also due to a local process, where the core regions have a higher \CII\ deficit than the regions further out from the core. 

\cite{smith17} suggest that the \CII\ deficit is related to the metallicity of the gas, where a low metallicity results in a high \LCII/\LFIR\ ratio.
Alternatively a recent study by \cite{munoz16} explored the possibility of \CII\ saturation.
This hypotheses had been proposed before but had not been investigated in detail \citep[e.g.,][]{stacey10, diaz-santos13, magdis14, gullberg15}. 
\cite{munoz16} show that the \CII\ emission can be thermally saturated at high temperatures. 
At gas temperatures $>92$\,K (the ground state temperature of \CII) the \CII\ cooling rate becomes constant, forcing the gas to cool through other channels (e.g., the [O{\small I}]63\,$\mu$m fine structure line). 
This means that the \CII\ emission line saturates, and a \CII\ deficit can therefore occur as a result of the further increase of the far-infrared luminosity. 
By considering the specific \CII\ luminosity (the \CII\ luminosity to \CII\ mass ratio, \LCII/$M_{\text{[CII]}}$) and specific far-infrared luminosity (the far-infrared luminosity to gas mass ratio, \LFIR/$M_{\text{gas}}$), \cite{munoz16} predict an \LCII/\LFIR\ relation dependent on the infrared surface density ($\Sigma_{\text{IR}}$) and fraction of the gas mass in ionized carbon ($f_{\text{[CII]}}=M_{\text{[CII]}}/M_{\text{gas}}$):
\begin{align}
\frac{L_{\text{[CII]}}}{L_{\text{FIR}}} \sim 2.2\times10^{-3}\frac{f_{\text{[CII]}}}{0.13}{\left(\frac{\Sigma_{\text{IR}}}{10^{11}\text{L}_{\odot}\,\text{kpc}^{-2}}\right)}^{-1/2}.
\end{align}
Here $f_{\text{[CII]}}$ is estimated to between 0.10 and 0.17, assuming a fixed CO(1--0) to \CII\ luminosity ratio, the CO to H$_2$ conversion factor ($\alpha_{\text{CO}}$), a gas density higher than the critical density of \CII\ ($n_{\text{crit}}=2.7\times10^3$\,cm$^{-3}$, e.g., \citealt{stacey10}) and a temperature higher than 92\,K. 
This fraction does not take into account that some of the carbon is in the neutral phase, meaning that the actual mass fraction is likely to be lower. 
The relation between the \CII\ to far-infrared luminosity ratio as a function of the star-formation rate surface density found by \cite{munoz16} adopt fixed values of $G=100$ and $n_{\text{gas}}=10^4$\,cm$^{-3}$ for the radiation field strength and the gas density. We plot the predicted trend line on Fig.~\ref{fig:sfrdens} for $f_{\text{[CII]}}$=0.03 and 0.14 which span the range of the data, suggesting that fraction of the total mass in ionised carbon is between 3--14\%.

However, other studies, for example \cite{diaz-santos17} and \cite{lagache17}, argue the \CII\ deficit arises from other factors. In particular, they emphasise the importance of a varying radiation field strength to gas density ratio ($G/n_{\text{gas}}$). \cite{diaz-santos17} fits a power law to local ULIRGs from GOALS (see Equation 3 in \citealt{diaz-santos17}) and find that the suppression of the \CII\ to far-infrared luminosity ratio at high far-infrared surface densities could be due to high $G/n_{\text{gas}}$ ratios. 
Using semi-analytical models \cite{lagache17} likewise suggest that the \CII\ to far-infrared deficit is correlated with the intensity of the interstellar radiation field.
This suggests that the star-formation rate surface density is dependent on the geometric of the photon dominated regions and the distribution of the gas and dust within it. 

We show the relation between the \CII\ to far-infrared luminosity ratio and the star-formation rate surface density with $f_{\text{[CII]}}=0.10-0.17$ in Fig.~\ref{fig:sfrdens} along with the power law relation determined by \cite{diaz-santos17}. 
Both the power law from \cite{diaz-santos17} and the model of \cite{munoz16} are in agreement with our observations. 
This means that it is not possible with the existing data to distinguish between the two models and determine if the \CII\ deficit can be explained by a saturation of the \CII\ emission at high temperatures and densities in the dense core regions of the SMGs or a high ratio of radiation field strength to gas density. 
We also note that contributions from other local processes (e.g.,\ \CII\ self-absorption, dust extinction, and dust grain charge), may also play a role in the deficit \citep{smith17}. 

\section{Conclusions} \label{sec:con}
\noindent
We present deep high-resolution (0\farcs03) ALMA Band 7 observations of the dust continuum and the redshifted \CII\ $158\,\mu$m emission line in four SMGs from the ALESS and AS2UDS surveys at $z\sim4.4-4.8$. 
These observations resolve the dust and gas distribution on $\sim200$\,pc scales and reveal a range of morphologies, ranging from compact and smooth to extended and apparently clumpy. 
\\\\
$\bullet$ By determining the amplitudes as function of the $uv$-distance for both the continuum and \CII\ emission, we conclude that the \CII\ emission is more extended than the rest-frame 160\,$\mu$m dust continuum emission by a factor of $2.1\pm0.4$.
This behavior is also seen in a sample of high-redshift galaxies from the literature, where we find a mean ratio of the \CII\ to rest-frame 160\,$\mu$m dust size of  $1.6\pm0.4$.
\\\\
$\bullet$ Three of our four SMGs show smooth continuum morphologies at sub-kpc resolution, the fourth, \Afourseven\ appears clumpy at 200\,pc resolution. To determine whether the apparent clumps in \Afourseven\ are real we compare with simulated observations of smooth exponential disks. 
These comparisons show that smooth morphologies can appear clumpy if observed at high resolution, even in deep observations such as those used here. 
We conclude that it is not possible to rule out the hypothesis that all of our SMGs are smooth exponential disks. 
Deeper observations are required to further test this. 
\\\\
$\bullet$ By comparing the \LCII/\LFIR\ ratio as a function of the star-formation rate surface density for the core with a diameter of $\sim2$\,kpc of our SMGs to that in the lower-density outskirts, we conclude that the \CII\ deficit is likely to be due to local processes, which is in agreement with the conclusion of \cite{smith17}. 
Comparing the observed relation between the \LCII/\LFIR\ ratio and the star-formation rate surface density, we find trends which agree with both the relation derived by \cite{munoz16} based on a thermal saturation of the \CII\ emission and \cite{diaz-santos17} based on high radiation field strength to gas density ratios. 
It is not possible therefore at this stage to determine which of these models best explains the \CII\ deficit.

Deeper \CII\ observations with $uv$-coverage at both long, medium and short baselines are necessary to test the hypothesis that SMGs having smooth or clumpy structures and to establish what physical processes drive the \CII\ deficit.

\acknowledgments
\noindent
We thank the anonymous referee for her/his their helpful thorough reading of the manuscript, and suggestions that improved the paper.
BG, EAC, and IRS acknowledge support from the ERC Advanced Programme DUSTYGAL (321334) and STFC (ST/P0000541/1). 
IRS also acknowledge support from a Royal Society Wolfson Merit Award.
JLW acknowledges support from an European Union COFUND/Durham Junior Research Fellowship (EU grant agreement number 609412) and from STFC (via an Ernest Rutherford Fellowship: ST/P004784/1, and additionally ST/P0000541/1)
EI acknowledges partial support from FONDECYT through grant N$^\circ$\,1171710.
This paper makes use of the following ALMA data: ADS/JAO.ALMA\#2015.1.00456.S ALMA is a partnership of ESO (representing its member states), NSF (USA) and NINS (Japan), together with NRC (Canada), MOST and ASIAA (Taiwan), and KASI (Republic of Korea), in cooperation with the Republic of Chile. The Joint ALMA Observatory is operated by ESO, AUI/NRAO and NAOJ.

\vspace{6mm}

\bibliography{ALESS}

\begin{thebibliography}{}
\expandafter\ifx\csname natexlab\endcsname\relax\def\natexlab#1{#1}\fi

\bibitem[{{ALMA Partnership} {et~al.}(2015){ALMA Partnership}, {Vlahakis},
  {Hunter}, {Hodge}, {P{\'e}rez}, {Andreani}, {Brogan}, {Cox}, {Martin},
  {Zwaan}, {Matsushita}, {Dent}, {Impellizzeri}, {Fomalont}, {Asaki},
  {Barkats}, {Hills}, {Hirota}, {Kneissl}, {Liuzzo}, {Lucas}, {Marcelino},
  {Nakanishi}, {Phillips}, {Richards}, {Toledo}, {Aladro}, {Broguiere},
  {Cortes}, {Cortes}, {Espada}, {Galarza}, {Garcia-Appadoo}, {Guzman-Ramirez},
  {Hales}, {Humphreys}, {Jung}, {Kameno}, {Laing}, {Leon}, {Marconi},
  {Mignano}, {Nikolic}, {Nyman}, {Radiszcz}, {Remijan}, {Rod{\'o}n}, {Sawada},
  {Takahashi}, {Tilanus}, {Vila Vilaro}, {Watson}, {Wiklind}, {Ao}, {Di
  Francesco}, {Hatsukade}, {Hatziminaoglou}, {Mangum}, {Matsuda}, {van Kampen},
  {Wootten}, {de Gregorio-Monsalvo}, {Dumas}, {Francke}, {Gallardo}, {Garcia},
  {Gonzalez}, {Hill}, {Iono}, {Kaminski}, {Karim}, {Krips}, {Kurono},
  {Lonsdale}, {Lopez}, {Morales}, {Plarre}, {Videla}, {Villard}, {Hibbard}, \&
  {Tatematsu}}]{almapartnership15}
{ALMA Partnership}, {Vlahakis}, C., {Hunter}, T.~R., {et~al.} 2015, \apjl, 808,
  L4

\bibitem[{{Armus} {et~al.}(2009){Armus}, {Mazzarella}, {Evans}, {Surace},
  {Sanders}, {Iwasawa}, {Frayer}, {Howell}, {Chan}, {Petric}, {Vavilkin},
  {Kim}, {Haan}, {Inami}, {Murphy}, {Appleton}, {Barnes}, {Bothun}, {Bridge},
  {Charmandaris}, {Jensen}, {Kewley}, {Lord}, {Madore}, {Marshall},
  {Melbourne}, {Rich}, {Satyapal}, {Schulz}, {Spoon}, {Sturm}, {U}, {Veilleux},
  \& {Xu}}]{armus09}
{Armus}, L., {Mazzarella}, J.~M., {Evans}, A.~S., {et~al.} 2009, \pasp, 121,
  559

\bibitem[{{Bothwell} {et~al.}(2013){Bothwell}, {Smail}, {Chapman}, {Genzel},
  {Ivison}, {Tacconi}, {Alaghband-Zadeh}, {Bertoldi}, {Blain}, {Casey}, {Cox},
  {Greve}, {Lutz}, {Neri}, {Omont}, \& {Swinbank}}]{bothwell13}
{Bothwell}, M.~S., {Smail}, I., {Chapman}, S.~C., {et~al.} 2013, \mnras, 429,
  3047

\bibitem[{{Bournaud} {et~al.}(2014){Bournaud}, {Perret}, {Renaud}, {Dekel},
  {Elmegreen}, {Elmegreen}, {Teyssier}, {Amram}, {Daddi}, {Duc}, {Elbaz},
  {Epinat}, {Gabor}, {Juneau}, {Kraljic}, \& {Le Floch'}}]{bournaud14}
{Bournaud}, F., {Perret}, V., {Renaud}, F., {et~al.} 2014, \apj, 780, 57

\bibitem[{{Brauher} {et~al.}(2008){Brauher}, {Dale}, \& {Helou}}]{brauher08}
{Brauher}, J.~R., {Dale}, D.~A., \& {Helou}, G. 2008, \apjs, 178, 280

\bibitem[{{Capak} {et~al.}(2015){Capak}, {Carilli}, {Jones}, {Casey},
  {Riechers}, {Sheth}, {Carollo}, {Ilbert}, {Karim}, {Lefevre}, {Lilly},
  {Scoville}, {Smolcic}, \& {Yan}}]{capak15}
{Capak}, P.~L., {Carilli}, C., {Jones}, G., {et~al.} 2015, \nat, 522, 455

\bibitem[{{Carniani} {et~al.}(2013){Carniani}, {Marconi}, {Biggs}, {Cresci},
  {Cupani}, {D'Odorico}, {Humphreys}, {Maiolino}, {Mannucci}, {Molaro},
  {Nagao}, {Testi}, \& {Zwaan}}]{carniani13}
{Carniani}, S., {Marconi}, A., {Biggs}, A., {et~al.} 2013, \aap, 559, A29

\bibitem[{{Casey} {et~al.}(2014){Casey}, {Narayanan}, \& {Cooray}}]{casey14}
{Casey}, C.~M., {Narayanan}, D., \& {Cooray}, A. 2014, \physrep, 541, 45

\bibitem[{{Chapman} {et~al.}(2005){Chapman}, {Blain}, {Smail}, \&
  {Ivison}}]{chapman05}
{Chapman}, S.~C., {Blain}, A.~W., {Smail}, I., \& {Ivison}, R.~J. 2005, \apj,
  622, 772

\bibitem[{{Chapman} {et~al.}(2004){Chapman}, {Smail}, {Blain}, \&
  {Ivison}}]{chapman04}
{Chapman}, S.~C., {Smail}, I., {Blain}, A.~W., \& {Ivison}, R.~J. 2004, \apj,
  614, 671

\bibitem[{{Chen} {et~al.}(2015){Chen}, {Smail}, {Swinbank}, {Simpson}, {Ma},
  {Alexander}, {Biggs}, {Brandt}, {Chapman}, {Coppin}, {Danielson},
  {Dannerbauer}, {Edge}, {Greve}, {Ivison}, {Karim}, {Menten}, {Schinnerer},
  {Walter}, {Wardlow}, {Wei{\ss}}, \& {van der Werf}}]{chen15}
{Chen}, C.-C., {Smail}, I., {Swinbank}, A.~M., {et~al.} 2015, \apj, 799, 194

\bibitem[{{Clements} \& {Baker}(1996)}]{clements96}
{Clements}, D.~L., \& {Baker}, A.~C. 1996, \aap, 314, L5

\bibitem[{{Coppin} {et~al.}(2009){Coppin}, {Smail}, {Alexander}, {Weiss},
  {Walter}, {Swinbank}, {Greve}, {Kovacs}, {De Breuck}, {Dickinson}, {Ibar},
  {Ivison}, {Reddy}, {Spinrad}, {Stern}, {Brandt}, {Chapman}, {Dannerbauer},
  {van Dokkum}, {Dunlop}, {Frayer}, {Gawiser}, {Geach}, {Huynh}, {Knudsen},
  {Koekemoer}, {Lehmer}, {Menten}, {Papovich}, {Rix}, {Schinnerer}, {Wardlow},
  \& {van der Werf}}]{coppin09}
{Coppin}, K.~E.~K., {Smail}, I., {Alexander}, D.~M., {et~al.} 2009, \mnras,
  395, 1905

\bibitem[{{Cowley} {et~al.}(2017){Cowley}, {B{\'e}thermin}, {Lagos}, {Lacey},
  {Baugh}, \& {Cole}}]{cowley17}
{Cowley}, W.~I., {B{\'e}thermin}, M., {Lagos}, C.~d.~P., {et~al.} 2017, \mnras,
  467, 1231

\bibitem[{{Croxall} {et~al.}(2012){Croxall}, {Smith}, {Wolfire}, {Roussel},
  {Sandstrom}, {Draine}, {Aniano}, {Dale}, {Armus}, {Beir{\~a}o}, {Helou},
  {Bolatto}, {Appleton}, {Brandl}, {Calzetti}, {Crocker}, {Galametz}, {Groves},
  {Hao}, {Hunt}, {Johnson}, {Kennicutt}, {Koda}, {Krause}, {Li}, {Meidt},
  {Murphy}, {Rahman}, {Rix}, {Sauvage}, {Schinnerer}, {Walter}, \&
  {Wilson}}]{croxall12}
{Croxall}, K.~V., {Smith}, J.~D., {Wolfire}, M.~G., {et~al.} 2012, \apj, 747,
  81

\bibitem[{{Croxall} {et~al.}(2017){Croxall}, {Smith}, {Pellegrini}, {Groves},
  {Bolatto}, {Herrera-Camus}, {Sandstrom}, {Draine}, {Wolfire}, {Armus},
  {Boquien}, {Brandl}, {Dale}, {Galametz}, {Hunt}, {Kennicutt}, {Kreckel},
  {Rigopoulou}, {van der Werf}, \& {Wilson}}]{croxall17}
{Croxall}, K.~V., {Smith}, J.~D., {Pellegrini}, E., {et~al.} 2017, \apj, 845,
  96

\bibitem[{{da Cunha} {et~al.}(2013){da Cunha}, {Groves}, {Walter}, {Decarli},
  {Weiss}, {Bertoldi}, {Carilli}, {Daddi}, {Elbaz}, {Ivison}, {Maiolino},
  {Riechers}, {Rix}, {Sargent}, \& {Smail}}]{dacunha13}
{da Cunha}, E., {Groves}, B., {Walter}, F., {et~al.} 2013, \apj, 766, 13

\bibitem[{{Dale} {et~al.}(2016){Dale}, {Beltz-Mohrmann}, {Egan}, {Hatlestad},
  {Herzog}, {Leung}, {McLane}, {Phenicie}, {Roberts}, {Barnes}, {Boquien},
  {Calzetti}, {Cook}, {Kobulnicky}, {Staudaher}, \& {van Zee}}]{dale16}
{Dale}, D.~A., {Beltz-Mohrmann}, G.~D., {Egan}, A.~A., {et~al.} 2016, \aj, 151,
  4

\bibitem[{{De Breuck} {et~al.}(2014){De Breuck}, {Williams}, {Swinbank},
  {Caselli}, {Coppin}, {Davis}, {Maiolino}, {Nagao}, {Smail}, {Walter},
  {Wei{\ss}}, \& {Zwaan}}]{debreuck14}
{De Breuck}, C., {Williams}, R.~J., {Swinbank}, M., {et~al.} 2014, The
  Messenger, 156, 38

\bibitem[{{Dekel} {et~al.}(2009){Dekel}, {Sari}, \& {Ceverino}}]{dekel09}
{Dekel}, A., {Sari}, R., \& {Ceverino}, D. 2009, \apj, 703, 785

\bibitem[{{D{\'{\i}}az-Santos} {et~al.}(2013){D{\'{\i}}az-Santos}, {Armus},
  {Charmandaris}, {Stierwalt}, {Murphy}, {Haan}, {Inami}, {Malhotra},
  {Meijerink}, {Stacey}, {Petric}, {Evans}, {Veilleux}, {van der Werf}, {Lord},
  {Lu}, {Howell}, {Appleton}, {Mazzarella}, {Surace}, {Xu}, {Schulz},
  {Sanders}, {Bridge}, {Chan}, {Frayer}, {Iwasawa}, {Melbourne}, \&
  {Sturm}}]{diaz-santos13}
{D{\'{\i}}az-Santos}, T., {Armus}, L., {Charmandaris}, V., {et~al.} 2013, \apj,
  774, 68

\bibitem[{{D{\'{\i}}az-Santos} {et~al.}(2016){D{\'{\i}}az-Santos}, {Assef},
  {Blain}, {Tsai}, {Aravena}, {Eisenhardt}, {Wu}, {Stern}, \&
  {Bridge}}]{diaz-santos16}
{D{\'{\i}}az-Santos}, T., {Assef}, R.~J., {Blain}, A.~W., {et~al.} 2016, \apjl,
  816, L6

\bibitem[{{D{\'{\i}}az-Santos} {et~al.}(2017){D{\'{\i}}az-Santos}, {Armus},
  {Charmandaris}, {Lu}, {Stierwalt}, {Stacey}, {Malhotra}, {van der Werf},
  {Howell}, {Privon}, {Mazzarella}, {Goldsmith}, {Murphy}, {Barcos-Mu{\~n}oz},
  {Linden}, {Inami}, {Larson}, {Evans}, {Appleton}, {Iwasawa}, {Lord},
  {Sanders}, \& {Surace}}]{diaz-santos17}
{D{\'{\i}}az-Santos}, T., {Armus}, L., {Charmandaris}, V., {et~al.} 2017, \apj,
  846, 32

\bibitem[{{Elmegreen} {et~al.}(2009){Elmegreen}, {Elmegreen}, {Marcus},
  {Shahinyan}, {Yau}, \& {Petersen}}]{elmegreen09}
{Elmegreen}, D.~M., {Elmegreen}, B.~G., {Marcus}, M.~T., {et~al.} 2009, \apj,
  701, 306

\bibitem[{{Engel} {et~al.}(2010){Engel}, {Tacconi}, {Davies}, {Neri}, {Smail},
  {Chapman}, {Genzel}, {Cox}, {Greve}, {Ivison}, {Blain}, {Bertoldi}, \&
  {Omont}}]{engel10}
{Engel}, H., {Tacconi}, L.~J., {Davies}, R.~I., {et~al.} 2010, \apj, 724, 233

\bibitem[{{Faisst} {et~al.}(2017){Faisst}, {Capak}, {Yan}, {Pavesi},
  {Riechers}, {Bari{\v s}i{\'c}}, {Cooke}, {Kartaltepe}, \&
  {Masters}}]{faisst17}
{Faisst}, A.~L., {Capak}, P.~L., {Yan}, L., {et~al.} 2017, \apj, 847, 21

\bibitem[{{Farrah} {et~al.}(2001){Farrah}, {Rowan-Robinson}, {Oliver},
  {Serjeant}, {Borne}, {Lawrence}, {Lucas}, {Bushouse}, \& {Colina}}]{farrah01}
{Farrah}, D., {Rowan-Robinson}, M., {Oliver}, S., {et~al.} 2001, \mnras, 326,
  1333

\bibitem[{{Farrah} {et~al.}(2013){Farrah}, {Lebouteiller}, {Spoon},
  {Bernard-Salas}, {Pearson}, {Rigopoulou}, {Smith}, {Gonz{\'a}lez-Alfonso},
  {Clements}, {Efstathiou}, {Cormier}, {Afonso}, {Petty}, {Harris}, {Hurley},
  {Borys}, {Verma}, {Cooray}, \& {Salvatelli}}]{farrah13}
{Farrah}, D., {Lebouteiller}, V., {Spoon}, H.~W.~W., {et~al.} 2013, \apj, 776,
  38

\bibitem[{{F{\"o}rster Schreiber} {et~al.}(2011){F{\"o}rster Schreiber},
  {Shapley}, {Genzel}, {Bouch{\'e}}, {Cresci}, {Davies}, {Erb}, {Genel},
  {Lutz}, {Newman}, {Shapiro}, {Steidel}, {Sternberg}, \&
  {Tacconi}}]{forster-schreiber11}
{F{\"o}rster Schreiber}, N.~M., {Shapley}, A.~E., {Genzel}, R., {et~al.} 2011,
  \apj, 739, 45

\bibitem[{{Geach} {et~al.}(2017){Geach}, {Dunlop}, {Halpern}, {Smail}, {van der
  Werf}, {Alexander}, {Almaini}, {Aretxaga}, {Arumugam}, {Asboth}, {Banerji},
  {Beanlands}, {Best}, {Blain}, {Birkinshaw}, {Chapin}, {Chapman}, {Chen},
  {Chrysostomou}, {Clarke}, {Clements}, {Conselice}, {Coppin}, {Cowley},
  {Danielson}, {Eales}, {Edge}, {Farrah}, {Gibb}, {Harrison}, {Hine}, {Hughes},
  {Ivison}, {Jarvis}, {Jenness}, {Jones}, {Karim}, {Koprowski}, {Knudsen},
  {Lacey}, {Mackenzie}, {Marsden}, {McAlpine}, {McMahon}, {Meijerink},
  {Micha{\l}owski}, {Oliver}, {Page}, {Peacock}, {Rigopoulou}, {Robson},
  {Roseboom}, {Rotermund}, {Scott}, {Serjeant}, {Simpson}, {Simpson}, {Smith},
  {Spaans}, {Stanley}, {Stevens}, {Swinbank}, {Targett}, {Thomson}, {Valiante},
  {Wake}, {Webb}, {Willott}, {Zavala}, \& {Zemcov}}]{geach17}
{Geach}, J.~E., {Dunlop}, J.~S., {Halpern}, M., {et~al.} 2017, \mnras, 465,
  1789

\bibitem[{{Genzel} {et~al.}(2012){Genzel}, {Tacconi}, {Combes}, {Bolatto},
  {Neri}, {Sternberg}, {Cooper}, {Bouch{\'e}}, {Bournaud}, {Burkert},
  {Comerford}, {Cox}, {Davis}, {F{\"o}rster Schreiber}, {Garcia-Burillo},
  {Gracia-Carpio}, {Lutz}, {Naab}, {Newman}, {Saintonge}, {Shapiro}, {Shapley},
  \& {Weiner}}]{genzel12}
{Genzel}, R., {Tacconi}, L.~J., {Combes}, F., {et~al.} 2012, \apj, 746, 69

\bibitem[{{Graci{\'a}-Carpio} {et~al.}(2011){Graci{\'a}-Carpio}, {Sturm},
  {Hailey-Dunsheath}, {Fischer}, {Contursi}, {Poglitsch}, {Genzel},
  {Gonz{\'a}lez-Alfonso}, {Sternberg}, {Verma}, {Christopher}, {Davies},
  {Feuchtgruber}, {de Jong}, {Lutz}, \& {Tacconi}}]{gracia_carpio11}
{Graci{\'a}-Carpio}, J., {Sturm}, E., {Hailey-Dunsheath}, S., {et~al.} 2011,
  \apjl, 728, L7

\bibitem[{{Gullberg} {et~al.}(2015){Gullberg}, {De Breuck}, {Vieira},
  {Wei{\ss}}, {Aguirre}, {Aravena}, {B{\'e}thermin}, {Bradford}, {Bothwell},
  {Carlstrom}, {Chapman}, {Fassnacht}, {Gonzalez}, {Greve}, {Hezaveh},
  {Holzapfel}, {Husband}, {Ma}, {Malkan}, {Marrone}, {Menten}, {Murphy},
  {Reichardt}, {Spilker}, {Stark}, {Strandet}, \& {Welikala}}]{gullberg15}
{Gullberg}, B., {De Breuck}, C., {Vieira}, J.~D., {et~al.} 2015, \mnras, 449,
  2883

\bibitem[{{Hailey-Dunsheath} {et~al.}(2010){Hailey-Dunsheath}, {Nikola},
  {Stacey}, {Oberst}, {Parshley}, {Benford}, {Staguhn}, \&
  {Tucker}}]{hailey-dunsheath10}
{Hailey-Dunsheath}, S., {Nikola}, T., {Stacey}, G.~J., {et~al.} 2010, \apjl,
  714, L162

\bibitem[{{Hatsukade} {et~al.}(2015){Hatsukade}, {Tamura}, {Iono}, {Matsuda},
  {Hayashi}, \& {Oguri}}]{hatsukade15}
{Hatsukade}, B., {Tamura}, Y., {Iono}, D., {et~al.} 2015, \pasj, 67, 93

\bibitem[{{Hayward} {et~al.}(2011){Hayward}, {Kere{\v s}}, {Jonsson},
  {Narayanan}, {Cox}, \& {Hernquist}}]{hayward11}
{Hayward}, C.~C., {Kere{\v s}}, D., {Jonsson}, P., {et~al.} 2011, \apj, 743,
  159

\bibitem[{{Hill} {et~al.}(2005){Hill}, {Burton}, {Minier}, {Thompson}, {Walsh},
  {Hunt-Cunningham}, \& {Garay}}]{hill05}
{Hill}, T., {Burton}, M.~G., {Minier}, V., {et~al.} 2005, \mnras, 363, 405

\bibitem[{{Hodge} {et~al.}(2013){Hodge}, {Karim}, {Smail}, {Swinbank},
  {Walter}, {Biggs}, {Ivison}, {Weiss}, {Alexander}, {Bertoldi}, {Brandt},
  {Chapman}, {Coppin}, {Cox}, {Danielson}, {Dannerbauer}, {De Breuck},
  {Decarli}, {Edge}, {Greve}, {Knudsen}, {Menten}, {Rix}, {Schinnerer},
  {Simpson}, {Wardlow}, \& {van der Werf}}]{hodge13}
{Hodge}, J.~A., {Karim}, A., {Smail}, I., {et~al.} 2013, \apj, 768, 91

\bibitem[{{Hodge} {et~al.}(2016){Hodge}, {Swinbank}, {Simpson}, {Smail},
  {Walter}, {Alexander}, {Bertoldi}, {Biggs}, {Brandt}, {Chapman}, {Chen},
  {Coppin}, {Cox}, {Dannerbauer}, {Edge}, {Greve}, {Ivison}, {Karim},
  {Knudsen}, {Menten}, {Rix}, {Schinnerer}, {Wardlow}, {Weiss}, \& {van der
  Werf}}]{hodge16}
{Hodge}, J.~A., {Swinbank}, A.~M., {Simpson}, J.~M., {et~al.} 2016, \apj, 833,
  103

\bibitem[{{Ikarashi} {et~al.}(2015){Ikarashi}, {Ivison}, {Caputi}, {Aretxaga},
  {Dunlop}, {Hatsukade}, {Hughes}, {Iono}, {Izumi}, {Kawabe}, {Kohno}, {Lagos},
  {Motohara}, {Nakanishi}, {Ohta}, {Tamura}, {Umehata}, {Wilson}, {Yabe}, \&
  {Yun}}]{ikarashi15}
{Ikarashi}, S., {Ivison}, R.~J., {Caputi}, K.~I., {et~al.} 2015, \apj, 810, 133

\bibitem[{{Iono} {et~al.}(2016){Iono}, {Yun}, {Aretxaga}, {Hatsukade},
  {Hughes}, {Ikarashi}, {Izumi}, {Kawabe}, {Kohno}, {Lee}, {Matsuda},
  {Nakanishi}, {Saito}, {Tamura}, {Ueda}, {Umehata}, {Wilson}, {Michiyama}, \&
  {Ando}}]{iono16}
{Iono}, D., {Yun}, M.~S., {Aretxaga}, I., {et~al.} 2016, \apjl, 829, L10

\bibitem[{{Ivison} {et~al.}(2010){Ivison}, {Swinbank}, {Swinyard}, {Smail},
  {Pearson}, {Rigopoulou}, {Polehampton}, {Baluteau}, {Barlow}, {Blain},
  {Bock}, {Clements}, {Coppin}, {Cooray}, {Danielson}, {Dwek}, {Edge},
  {Franceschini}, {Fulton}, {Glenn}, {Griffin}, {Isaak}, {Leeks}, {Lim},
  {Naylor}, {Oliver}, {Page}, {P{\'e}rez Fournon}, {Rowan-Robinson}, {Savini},
  {Scott}, {Spencer}, {Valtchanov}, {Vigroux}, \& {Wright}}]{ivison10}
{Ivison}, R.~J., {Swinbank}, A.~M., {Swinyard}, B., {et~al.} 2010, \aap, 518,
  L35

\bibitem[{{Ivison} {et~al.}(2012){Ivison}, {Smail}, {Amblard}, {Arumugam}, {De
  Breuck}, {Emonts}, {Feain}, {Greve}, {Haas}, {Ibar}, {Jarvis}, {Kov{\'a}cs},
  {Lehnert}, {Nesvadba}, {R{\"o}ttgering}, {Seymour}, \&
  {Wylezalek}}]{ivison12}
{Ivison}, R.~J., {Smail}, I., {Amblard}, A., {et~al.} 2012, \mnras, 425, 1320

\bibitem[{{James} {et~al.}(2002){James}, {Dunne}, {Eales}, \&
  {Edmunds}}]{james02}
{James}, A., {Dunne}, L., {Eales}, S., \& {Edmunds}, M.~G. 2002, \mnras, 335,
  753

\bibitem[{{Jones} {et~al.}(2017){Jones}, {Willott}, {Carilli}, {Ferrara},
  {Wang}, \& {Wagg}}]{jones17}
{Jones}, G.~C., {Willott}, C.~J., {Carilli}, C.~L., {et~al.} 2017, \apj, 845,
  175

\bibitem[{{Kennicutt} {et~al.}(2011){Kennicutt}, {Calzetti}, {Aniano},
  {Appleton}, {Armus}, {Beir{\~a}o}, {Bolatto}, {Brandl}, {Crocker}, {Croxall},
  {Dale}, {Donovan Meyer}, {Draine}, {Engelbracht}, {Galametz}, {Gordon},
  {Groves}, {Hao}, {Helou}, {Hinz}, {Hunt}, {Johnson}, {Koda}, {Krause},
  {Leroy}, {Li}, {Meidt}, {Montiel}, {Murphy}, {Rahman}, {Rix}, {Roussel},
  {Sandstrom}, {Sauvage}, {Schinnerer}, {Skibba}, {Smith}, {Srinivasan},
  {Vigroux}, {Walter}, {Wilson}, {Wolfire}, \& {Zibetti}}]{kennicutt11}
{Kennicutt}, R.~C., {Calzetti}, D., {Aniano}, G., {et~al.} 2011, \pasp, 123,
  1347

\bibitem[{{Kennicutt}(1998)}]{kennicutt98}
{Kennicutt}, Jr., R.~C. 1998, \araa, 36, 189

\bibitem[{{Kennicutt} \& {Chu}(1988)}]{kennicutt88}
{Kennicutt}, Jr., R.~C., \& {Chu}, Y.-H. 1988, \aj, 95, 720

\bibitem[{{Kimball} {et~al.}(2015){Kimball}, {Lacy}, {Lonsdale}, \&
  {Macquart}}]{kimball15}
{Kimball}, A.~E., {Lacy}, M., {Lonsdale}, C.~J., \& {Macquart}, J.-P. 2015,
  \mnras, 452, 88

\bibitem[{{Lagache} {et~al.}(2017){Lagache}, {Cousin}, \&
  {Chatzikos}}]{lagache17}
{Lagache}, G., {Cousin}, M., \& {Chatzikos}, M. 2017, ArXiv e-prints,
  arXiv:1711.00798

\bibitem[{{Law} {et~al.}(2009){Law}, {Steidel}, {Erb}, {Larkin}, {Pettini},
  {Shapley}, \& {Wright}}]{law09}
{Law}, D.~R., {Steidel}, C.~C., {Erb}, D.~K., {et~al.} 2009, \apj, 697, 2057

\bibitem[{{Livermore} {et~al.}(2012){Livermore}, {Jones}, {Richard}, {Bower},
  {Ellis}, {Swinbank}, {Rigby}, {Smail}, {Arribas}, {Rodriguez Zaurin},
  {Colina}, {Ebeling}, \& {Crain}}]{livermore12}
{Livermore}, R.~C., {Jones}, T., {Richard}, J., {et~al.} 2012, \mnras, 427, 688

\bibitem[{{Lord} {et~al.}(1996){Lord}, {Malhotra}, {Helou}, {Beichman}, {Lu},
  {Lim}, {Hollenbach}, {Rubin}, {Thronson}, {Stacey}, {Dinerstein}, {Werner},
  {Hunter}, \& {Lo}}]{lord96}
{Lord}, S., {Malhotra}, S., {Helou}, G., {et~al.} 1996, in Bulletin of the
  American Astronomical Society, Vol.~28, American Astronomical Society Meeting
  Abstracts \#188, 929

\bibitem[{{Luhman} {et~al.}(2003){Luhman}, {Satyapal}, {Fischer}, {Wolfire},
  {Sturm}, {Dudley}, {Lutz}, \& {Genzel}}]{luhman03}
{Luhman}, M.~L., {Satyapal}, S., {Fischer}, J., {et~al.} 2003, \apj, 594, 758

\bibitem[{{Luhman} {et~al.}(1998){Luhman}, {Satyapal}, {Fischer}, {Wolfire},
  {Cox}, {Lord}, {Smith}, {Stacey}, \& {Unger}}]{luhman98}
---. 1998, \apjl, 504, L11

\bibitem[{{Lutz} {et~al.}(2016{\natexlab{a}}){Lutz}, {Berta}, {Contursi},
  {F{\"o}rster Schreiber}, {Genzel}, {Graci{\'a}-Carpio}, {Herrera-Camus},
  {Netzer}, {Sturm}, {Tacconi}, {Tadaki}, \& {Veilleux}}]{spilker16}
{Lutz}, D., {Berta}, S., {Contursi}, A., {et~al.} 2016{\natexlab{a}}, \aap,
  591, A136

\bibitem[{{Lutz} {et~al.}(2016{\natexlab{b}}){Lutz}, {Berta}, {Contursi},
  {F{\"o}rster Schreiber}, {Genzel}, {Graci{\'a}-Carpio}, {Herrera-Camus},
  {Netzer}, {Sturm}, {Tacconi}, {Tadaki}, \& {Veilleux}}]{lutz16}
---. 2016{\natexlab{b}}, \aap, 591, A136

\bibitem[{{Madden} {et~al.}(1993){Madden}, {Geis}, {Genzel}, {Herrmann},
  {Jackson}, {Poglitsch}, {Stacey}, \& {Townes}}]{madden93}
{Madden}, S.~C., {Geis}, N., {Genzel}, R., {et~al.} 1993, \apj, 407, 579

\bibitem[{{Magdis} {et~al.}(2014){Magdis}, {Rigopoulou}, {Hopwood}, {Huang},
  {Farrah}, {Pearson}, {Alonso-Herrero}, {Bock}, {Clements}, {Cooray},
  {Griffin}, {Oliver}, {Perez Fournon}, {Riechers}, {Swinyard}, {Scott},
  {Thatte}, {Valtchanov}, \& {Vaccari}}]{magdis14}
{Magdis}, G.~E., {Rigopoulou}, D., {Hopwood}, R., {et~al.} 2014, \apj, 796, 63

\bibitem[{{Malhotra}(2001)}]{malhotra01}
{Malhotra}, S. 2001, in ESA Special Publication, Vol. 460, The Promise of the
  Herschel Space Observatory, ed. G.~L. {Pilbratt}, J.~{Cernicharo}, A.~M.
  {Heras}, T.~{Prusti}, \& R.~{Harris}, 155

\bibitem[{{Malhotra} {et~al.}(1997){Malhotra}, {Helou}, {Stacey}, {Hollenbach},
  {Lord}, {Beichman}, {Dinerstein}, {Hunter}, {Lo}, {Lu}, {Rubin},
  {Silbermann}, {Thronson}, \& {Werner}}]{malhotra97}
{Malhotra}, S., {Helou}, G., {Stacey}, G., {et~al.} 1997, \apjl, 491, L27

\bibitem[{{McMullin} {et~al.}(2007){McMullin}, {Waters}, {Schiebel}, {Young},
  \& {Golap}}]{mcmullin07}
{McMullin}, J.~P., {Waters}, B., {Schiebel}, D., {Young}, W., \& {Golap}, K.
  2007, in Astronomical Society of the Pacific Conference Series, Vol. 376,
  Astronomical Data Analysis Software and Systems XVI, ed. R.~A. {Shaw},
  F.~{Hill}, \& D.~J. {Bell}, 127

\bibitem[{{Mu{\~n}oz} \& {Oh}(2016)}]{munoz16}
{Mu{\~n}oz}, J.~A., \& {Oh}, S.~P. 2016, \mnras, 463, 2085

\bibitem[{{Nesvadba} {et~al.}(2016){Nesvadba}, {Kneissl}, {Ca{\~n}ameras},
  {Boone}, {Falgarone}, {Frye}, {Gerin}, {Koenig}, {Lagache}, {Le Floc'h},
  {Malhotra}, \& {Scott}}]{nesvadba16}
{Nesvadba}, N., {Kneissl}, R., {Ca{\~n}ameras}, R., {et~al.} 2016, \aap, 593,
  L2

\bibitem[{{Oteo} {et~al.}(2017){Oteo}, {Zwaan}, {Ivison}, {Smail}, \&
  {Biggs}}]{oteo17}
{Oteo}, I., {Zwaan}, M.~A., {Ivison}, R.~J., {Smail}, I., \& {Biggs}, A.~D.
  2017, \apj, 837, 182

\bibitem[{{Riechers} {et~al.}(2011){Riechers}, {Cooray}, {Omont}, {Neri},
  {Harris}, {Baker}, {Cox}, {Frayer}, {Carpenter}, {Auld}, {Aussel}, {Beelen},
  {Blundell}, {Bock}, {Brisbin}, {Burgarella}, {Chanial}, {Chapman},
  {Clements}, {Conley}, {Dowell}, {Eales}, {Farrah}, {Franceschini}, {Gavazzi},
  {Glenn}, {Griffin}, {Gurwell}, {Ivison}, {Kim}, {Krips}, {Mortier}, {Oliver},
  {Page}, {Papageorgiou}, {Pearson}, {P{\'e}rez-Fournon}, {Pohlen}, {Rawlings},
  {Raymond}, {Rodighiero}, {Roseboom}, {Rowan-Robinson}, {Scott}, {Seymour},
  {Smith}, {Symeonidis}, {Tugwell}, {Vaccari}, {Vieira}, {Vigroux}, {Wang},
  {Wardlow}, \& {Wright}}]{riechers11}
{Riechers}, D.~A., {Cooray}, A., {Omont}, A., {et~al.} 2011, \apjl, 733, L12

\bibitem[{{Sakamoto} {et~al.}(2008){Sakamoto}, {Wang}, {Wiedner}, {Wang},
  {Peck}, {Zhang}, {Petitpas}, {Ho}, \& {Wilner}}]{sakamoto08}
{Sakamoto}, K., {Wang}, J., {Wiedner}, M.~C., {et~al.} 2008, \apj, 684, 957

\bibitem[{{Sanders} \& {Mirabel}(1996)}]{sanders96}
{Sanders}, D.~B., \& {Mirabel}, I.~F. 1996, \araa, 34, 749

\bibitem[{{Shibuya} {et~al.}(2015){Shibuya}, {Ouchi}, \&
  {Harikane}}]{shibuya15}
{Shibuya}, T., {Ouchi}, M., \& {Harikane}, Y. 2015, \apjs, 219, 15

\bibitem[{{Simpson} {et~al.}(2015{\natexlab{a}}){Simpson}, {Smail}, {Swinbank},
  {Chapman}, {Geach}, {Ivison}, {Thomson}, {Aretxaga}, {Blain}, {Cowley},
  {Chen}, {Coppin}, {Dunlop}, {Edge}, {Farrah}, {Ibar}, {Karim}, {Knudsen},
  {Meijerink}, {Micha{\l}owski}, {Scott}, {Spaans}, \& {van der
  Werf}}]{simpson15b}
{Simpson}, J.~M., {Smail}, I., {Swinbank}, A.~M., {et~al.} 2015{\natexlab{a}},
  \apj, 807, 128

\bibitem[{{Simpson} {et~al.}(2015{\natexlab{b}}){Simpson}, {Smail}, {Swinbank},
  {Almaini}, {Blain}, {Bremer}, {Chapman}, {Chen}, {Conselice}, {Coppin},
  {Danielson}, {Dunlop}, {Edge}, {Farrah}, {Geach}, {Hartley}, {Ivison},
  {Karim}, {Lani}, {Ma}, {Meijerink}, {Micha{\l}owski}, {Mortlock}, {Scott},
  {Simpson}, {Spaans}, {Thomson}, {van Kampen}, \& {van der Werf}}]{simpson15a}
---. 2015{\natexlab{b}}, \apj, 799, 81

\bibitem[{{Simpson} {et~al.}(2017){Simpson}, {Smail}, {Swinbank}, {Ivison},
  {Dunlop}, {Geach}, {Almaini}, {Arumugam}, {Bremer}, {Chen}, {Conselice},
  {Coppin}, {Farrah}, {Ibar}, {Hartley}, {Ma}, {Micha{\l}owski}, {Scott},
  {Spaans}, {Thomson}, \& {van der Werf}}]{simpson17}
---. 2017, \apj, 839, 58

\bibitem[{{Smith} {et~al.}(2017){Smith}, {Croxall}, {Draine}, {De Looze},
  {Sandstrom}, {Armus}, {Beir{\~a}o}, {Bolatto}, {Boquien}, {Brandl},
  {Crocker}, {Dale}, {Galametz}, {Groves}, {Helou}, {Herrera-Camus}, {Hunt},
  {Kennicutt}, {Walter}, \& {Wolfire}}]{smith17}
{Smith}, J.~D.~T., {Croxall}, K., {Draine}, B., {et~al.} 2017, \apj, 834, 5

\bibitem[{{Stacey} {et~al.}(1991){Stacey}, {Geis}, {Genzel}, {Lugten},
  {Poglitsch}, {Sternberg}, \& {Townes}}]{stacey91}
{Stacey}, G.~J., {Geis}, N., {Genzel}, R., {et~al.} 1991, \apj, 373, 423

\bibitem[{{Stacey} {et~al.}(2010){Stacey}, {Hailey-Dunsheath}, {Ferkinhoff},
  {Nikola}, {Parshley}, {Benford}, {Staguhn}, \& {Fiolet}}]{stacey10}
{Stacey}, G.~J., {Hailey-Dunsheath}, S., {Ferkinhoff}, C., {et~al.} 2010, \apj,
  724, 957

\bibitem[{{Surace} {et~al.}(2001){Surace}, {Sanders}, \& {Evans}}]{surace01}
{Surace}, J.~A., {Sanders}, D.~B., \& {Evans}, A.~S. 2001, \aj, 122, 2791

\bibitem[{{Swinbank} {et~al.}(2010){Swinbank}, {Smail}, {Longmore}, {Harris},
  {Baker}, {De Breuck}, {Richard}, {Edge}, {Ivison}, {Blundell}, {Coppin},
  {Cox}, {Gurwell}, {Hainline}, {Krips}, {Lundgren}, {Neri}, {Siana},
  {Siringo}, {Stark}, {Wilner}, \& {Younger}}]{swinbank10}
{Swinbank}, A.~M., {Smail}, I., {Longmore}, S., {et~al.} 2010, \nat, 464, 733

\bibitem[{{Swinbank} {et~al.}(2012){Swinbank}, {Karim}, {Smail}, {Hodge},
  {Walter}, {Bertoldi}, {Biggs}, {de Breuck}, {Chapman}, {Coppin}, {Cox},
  {Danielson}, {Dannerbauer}, {Ivison}, {Greve}, {Knudsen}, {Menten},
  {Simpson}, {Schinnerer}, {Wardlow}, {Wei{\ss}}, \& {van der
  Werf}}]{swinbank12}
{Swinbank}, A.~M., {Karim}, A., {Smail}, I., {et~al.} 2012, \mnras, 427, 1066

\bibitem[{{Swinbank} {et~al.}(2014){Swinbank}, {Simpson}, {Smail}, {Harrison},
  {Hodge}, {Karim}, {Walter}, {Alexander}, {Brandt}, {de Breuck}, {da Cunha},
  {Chapman}, {Coppin}, {Danielson}, {Dannerbauer}, {Decarli}, {Greve},
  {Ivison}, {Knudsen}, {Lagos}, {Schinnerer}, {Thomson}, {Wardlow}, {Wei{\ss}},
  \& {van der Werf}}]{swinbank14}
{Swinbank}, A.~M., {Simpson}, J.~M., {Smail}, I., {et~al.} 2014, \mnras, 438,
  1267

\bibitem[{{Swinbank} {et~al.}(2015){Swinbank}, {Dye}, {Nightingale},
  {Furlanetto}, {Smail}, {Cooray}, {Dannerbauer}, {Dunne}, {Eales}, {Gavazzi},
  {Hunter}, {Ivison}, {Negrello}, {Oteo-Gomez}, {Smit}, {van der Werf}, \&
  {Vlahakis}}]{swinbank15}
{Swinbank}, A.~M., {Dye}, S., {Nightingale}, J.~W., {et~al.} 2015, \apjl, 806,
  L17

\bibitem[{{Tacconi} {et~al.}(2008){Tacconi}, {Genzel}, {Smail}, {Neri},
  {Chapman}, {Ivison}, {Blain}, {Cox}, {Omont}, {Bertoldi}, {Greve},
  {F{\"o}rster Schreiber}, {Genel}, {Lutz}, {Swinbank}, {Shapley}, {Erb},
  {Cimatti}, {Daddi}, \& {Baker}}]{tacconi08}
{Tacconi}, L.~J., {Genzel}, R., {Smail}, I., {et~al.} 2008, \apj, 680, 246

\bibitem[{{Tacconi} {et~al.}(2017){Tacconi}, {Genzel}, {Saintonge}, {Combes},
  {Garc{\'{\i}}a-Burillo}, {Neri}, {Bolatto}, {Contini}, {F{\"o}rster
  Schreiber}, {Lilly}, {Lutz}, {Wuyts}, {Accurso}, {Boissier}, {Boone},
  {Bouch{\'e}}, {Bournaud}, {Burkert}, {Carollo}, {Cooper}, {Cox}, {Feruglio},
  {Freundlich}, {Herrera-Camus}, {Juneau}, {Lippa}, {Naab}, {Renzini},
  {Salome}, {Sternberg}, {Tadaki}, {{\"U}bler}, {Walter}, {Weiner}, \&
  {Weiss}}]{tacconi17}
{Tacconi}, L.~J., {Genzel}, R., {Saintonge}, A., {et~al.} 2017, ArXiv e-prints,
  arXiv:1702.01140

\bibitem[{{Toomre}(1964)}]{toomre64}
{Toomre}, A. 1964, \apj, 139, 1217

\bibitem[{{Umehata} {et~al.}(2017){Umehata}, {Matsuda}, {Tamura}, {Kohno},
  {Smail}, {Ivison}, {Steidel}, {Chapman}, {Geach}, {Hayes}, {Nagao}, {Ao},
  {Kawabe}, {Yun}, {Hatsukade}, {Kubo}, {Kato}, {Saito}, {Ikarashi},
  {Nakanishi}, {Lee}, {Izumi}, {Mori}, \& {Ouchi}}]{umehata17}
{Umehata}, H., {Matsuda}, Y., {Tamura}, Y., {et~al.} 2017, \apjl, 834, L16

\bibitem[{{Valtchanov} {et~al.}(2011){Valtchanov}, {Virdee}, {Ivison},
  {Swinyard}, {van der Werf}, {Rigopoulou}, {da Cunha}, {Lupu}, {Benford},
  {Riechers}, {Smail}, {Jarvis}, {Pearson}, {Gomez}, {Hopwood}, {Altieri},
  {Birkinshaw}, {Coia}, {Conversi}, {Cooray}, {de Zotti}, {Dunne}, {Frayer},
  {Leeuw}, {Marston}, {Negrello}, {Portal}, {Scott}, {Thompson}, {Vaccari},
  {Baes}, {Clements}, {Micha{\l}owski}, {Dannerbauer}, {Serjeant}, {Auld},
  {Buttiglione}, {Cava}, {Dariush}, {Dye}, {Eales}, {Fritz}, {Ibar}, {Maddox},
  {Pascale}, {Pohlen}, {Rigby}, {Rodighiero}, {Smith}, {Temi}, {Carpenter},
  {Bolatto}, {Gurwell}, \& {Vieira}}]{valtchanov11}
{Valtchanov}, I., {Virdee}, J., {Ivison}, R.~J., {et~al.} 2011, \mnras, 415,
  3473

\bibitem[{{Veilleux}(2002)}]{veilleux02}
{Veilleux}, S. 2002, in Astronomical Society of the Pacific Conference Series,
  Vol. 254, Extragalactic Gas at Low Redshift, ed. J.~S. {Mulchaey} \& J.~T.
  {Stocke}, 313

\bibitem[{{Venemans} {et~al.}(2017){Venemans}, {Walter}, {Decarli},
  {Ba{\~n}ados}, {Hodge}, {Hewett}, {McMahon}, {Mortlock}, \&
  {Simpson}}]{venemans17}
{Venemans}, B.~P., {Walter}, F., {Decarli}, R., {et~al.} 2017, \apj, 837, 146

\bibitem[{{Wang} {et~al.}(2013){Wang}, {Wagg}, {Carilli}, {Walter}, {Lentati},
  {Fan}, {Riechers}, {Bertoldi}, {Narayanan}, {Strauss}, {Cox}, {Omont},
  {Menten}, {Knudsen}, {Neri}, \& {Jiang}}]{wang13}
{Wang}, R., {Wagg}, J., {Carilli}, C.~L., {et~al.} 2013, \apj, 773, 44

\bibitem[{{Wei{\ss}} {et~al.}(2009){Wei{\ss}}, {Kov{\'a}cs}, {Coppin}, {Greve},
  {Walter}, {Smail}, {Dunlop}, {Knudsen}, {Alexander}, {Bertoldi}, {Brandt},
  {Chapman}, {Cox}, {Dannerbauer}, {De Breuck}, {Gawiser}, {Ivison}, {Lutz},
  {Menten}, {Koekemoer}, {Kreysa}, {Kurczynski}, {Rix}, {Schinnerer}, \& {van
  der Werf}}]{weiss09}
{Wei{\ss}}, A., {Kov{\'a}cs}, A., {Coppin}, K., {et~al.} 2009, \apj, 707, 1201

\bibitem[{{Wei{\ss}} {et~al.}(2013){Wei{\ss}}, {De Breuck}, {Marrone},
  {Vieira}, {Aguirre}, {Aird}, {Aravena}, {Ashby}, {Bayliss}, {Benson},
  {B{\'e}thermin}, {Biggs}, {Bleem}, {Bock}, {Bothwell}, {Bradford}, {Brodwin},
  {Carlstrom}, {Chang}, {Chapman}, {Crawford}, {Crites}, {de Haan}, {Dobbs},
  {Downes}, {Fassnacht}, {George}, {Gladders}, {Gonzalez}, {Greve},
  {Halverson}, {Hezaveh}, {High}, {Holder}, {Holzapfel}, {Hoover}, {Hrubes},
  {Husband}, {Keisler}, {Lee}, {Leitch}, {Lueker}, {Luong-Van}, {Malkan},
  {McIntyre}, {McMahon}, {Mehl}, {Menten}, {Meyer}, {Murphy}, {Padin},
  {Plagge}, {Reichardt}, {Rest}, {Rosenman}, {Ruel}, {Ruhl}, {Schaffer},
  {Shirokoff}, {Spilker}, {Stalder}, {Staniszewski}, {Stark}, {Story},
  {Vanderlinde}, {Welikala}, \& {Williamson}}]{weiss13}
{Wei{\ss}}, A., {De Breuck}, C., {Marrone}, D.~P., {et~al.} 2013, \apj, 767, 88

\bibitem[{{Younger} {et~al.}(2008){Younger}, {Fazio}, {Wilner}, {Ashby},
  {Blundell}, {Gurwell}, {Huang}, {Iono}, {Peck}, {Petitpas}, {Scott},
  {Wilson}, \& {Yun}}]{younger08}
{Younger}, J.~D., {Fazio}, G.~G., {Wilner}, D.~J., {et~al.} 2008, \apj, 688, 59

\bibitem[{{Zhang} {et~al.}(2016){Zhang}, {Papadopoulos}, {Ivison}, {Galametz},
  {Smith}, \& {Xilouris}}]{zhang16}
{Zhang}, Z.-Y., {Papadopoulos}, P.~P., {Ivison}, R.~J., {et~al.} 2016, Royal
  Society Open Science, 3, 160025

\end{thebibliography}

\end{document}